\documentstyle[Dmath,11pt,epsf,ref]{article}
\def\simle{\mathrel{
   \rlap{\raise 0.511ex \hbox{$<$}}{\lower 0.511ex \hbox{$\sim$}}}}

\def\simge{\mathrel{%
   \rlap{\raise 0.511ex \hbox{$>$}}{\lower 0.511ex \hbox{$\sim$}}}}

\title{The Effects of the Tidal Force on Shear Instabilities 
in the Dust Layer of the Solar Nebula}

\author{Naoki Ishitsu$^1$ and Minoru Sekiya$^2$ \\
~\\
\small
E-mail: naoki.ishitsu@nao.ac.jp \\
\small
\it
$^1$Astronomical Data Analysis Center, National Astronomical Observatory,\\
\small
\it
Osawa 2-21-1, Mitaka, Tokyo 181-8588, Japan\\
\small
\it
$^2$Department of Earth and Planetary Sciences, Faculty of Sciences,\\ 
\small
\it
33 Kyushu University, Hakozaki, Fukuoka, 812-8581, Japan}

\date{}
\begin{document}
\maketitle
%\begin{titlepage}
%\begin{center}
%(Submitted 19 July, 2002; Revised 28 April, 2003)
%\end{center}

\begin{abstract}

The linear analysis of the instability due to 
vertical shear in the dust layer of the solar nebula is performed. 
The following assumptions are adopted throughout this paper:  
(1) The self-gravity of the dust layer is neglected.
(2) One fluid model is adopted,
where the dust aggregates have the same velocity with the gas 
due to strong coupling by the drag force.  
(3) The gas is incompressible.

The calculations with both the Coriolis and the tidal forces show 
that the tidal force has a stabilizing effect.  
The tidal force causes the radial shear  in the disk.  
This radial shear changes the wave number of the mode 
which is at first unstable, and the mode is eventually stabilized.  
Thus the behavior of the mode is divided into two stages:
(1) the first growth of the unstable mode 
which is similar to the results without the tidal force, and
(2) the subsequent stabilization due to an increase of the wave number 
by the radial shear.  
If the midplane dust/gas density ratio is smaller than 2, 
the stabilization occurs before the unstable mode grows largely. 
On the other hand, the mode grows faster by one hundred 
 orders of magnitude, if this ratio is larger than 20.

Because the critical density of the gravitational instability 
is a few hundreds times as large as the gas density, 
the hydrodynamic instability investigated  in this paper
grows largely before the onset of the gravitational instability.  
It is expected that 
the hydrodynamic  instability develops turbulence in the dust layer 
and the dust aggregates are stirred up to prevent 
from  settling further.  
The formation of planetesimals through the gravitational instabilities
is difficult to occur as long as the dust/gas surface density 
ratio is equal to that for the solar abundance.

On the other hand, the shear instability is suppressed and 
the planetesimal formation through the gravitational instability
may occur, if dust/gas surface density ratio is hundreds 
times as large as that for  the solar abundance.
\end{abstract}

{\it Key Words}: Planetary Formation; Solar Nebula,

\section{INTRODUCTION}
Planetesimal  formation in the solar nebula  
is one of unresolved issues in the theory of planetary formation.  
In the solar nebula, 
micron-sized dust particles which float uniformly at first 
are considered to stick each other 
by non-gravitational forces (e.g., the van der Waals force) 
to form dust aggregates.
If the solar nebula is laminar, 
the dust aggregates settle toward the midplane, 
and simultaneously they stick each other by collisions 
due to differential settling velocities.
Thus, a thin dust layer is formed around the midplane.  
The dust density around the midplane increases gradually 
due to the dust settling.  
When the dust self-gravity exceeds 
the tidal force of the central star, that is, the density exceeds
the critical density $\rho_c$,
the dust layer becomes gravitationally unstable and 
fragments into a lot of km-sized planetesimals
(Safronov 1969, Goldreich and Ward 1973,
 Coradini {\it et al.} 1981, Sekiya 1983).  
Formerly the above mentioned gravitational instability model was
considered to be the most promising process of planetesimal formation.  

However, if the solar nebula is turbulent, dust aggregates are stirred 
up, and it may be difficult for the dust density 
to exceed the critical density of the gravitational instability.  
Several causes of turbulence in the solar nebula
have been proposed: the thermal convection (Lin and Papaloizou 1980, 
Cabot {\it et al.} 1987a, b, Ruden {\it et al.} 1988), 
the hydrodynamic instability due to 
radial shear  in the disk (Papaloizou and Pringle 1984, 
Goldreich and Narayan 1985, Narayan {\it et al.} 1987, 
Nakagawa and Sekiya 1992, Sekiya {\it et al.} 1993), and 
the magneto-rotational instability (Balbus and Hawley 1991), and so on.  
This work concentrates on  investigating the instability 
due to vertical shear in the dust layer, 
which arises by the dust settling itself as explained in the following.
If the drag force did not work, dust aggregates would revolve 
with the Kepler velocity balancing the centrifugal force 
and the gravitational force of a central star. 
On the other hand, a gas fluid would revolve
 slightly slower than the Kepler velocity because only gas is partially
 supported by a gas pressure gradient. Thus, when gas and dust drag each other,
 their velocities depend on the dust to gas mass ratio and
 the frictional time. In particular, when the frictional time 
is much shorter than a Kepler period, namely that dust aggregate
 is small and/or fluffy, a dust aggregate and gas fluid move 
approximately with a  same velocity. 
 Then, the revolution velocity $\bar{v}$ in the co-ordinate system which moves
with the local Kepler velocity  $v_K$ is given by
\begin{equation}
\bar{v} = -\frac{\rho_g}{\rho} \eta v_K.
\label{eqn:introv}
\end{equation}
Here $\rho$ is the total fluid density defined by
\begin{equation}
	\rho = \rho_g + \rho_d,
\label{eqn:introrho}
\end{equation}
where $\rho_g$ is the gas density and $\rho_d$ is the dust density.
The non-dimensional parameter $\eta$ represents 
the effect of the radial pressure gradient:
\begin{equation}
	\eta = -\frac{r}{2\rho_gv_K^2}\frac{\partial P}{\partial r},
	\label{eqn:eta}
\end{equation}
where $r$ is the distance from the rotation axis of the disk and 
$P$ is the gas pressure (Adachi {\it et al.} 1976, Weidenschilling 1977,
Nakagawa {\it et al.} 1986, Sekiya 1998).
This vertical shear  may cause the shear instability, 
which may develop turbulence in the dust layer; 
thus dust aggregates may be stirred up 
and dust density may not exceed the critical density 
of the gravitational instability 
 (Weidenschilling 1980).  

Cuzzi {\it et al.} (1993) have performed numerical simulations 
of two fluids, gas and dust,  in order to examine 
whether the above mentioned turbulence is strong enough 
to prevent the gravitational instability of the dust layer.  
They assumed single-sized compact aggregates (10cm or 60cm).  
They obtained the quasi-equilibrium density distributions 
balancing the turbulent diffusion and the settling of aggregates.  
They showed that the density does not reach the critical density 
of the gravitational instability. 
Champney {\it et al.} (1995) extended this model to multi-fluids 
where different sized dust aggregates existed in the gas. 
Dobrovolskis {\it et al.} (1999) made calculations 
including the dissipation due to the friction 
between dust aggregates and the gas.  
The results of these papers are almost same as Cuzzi {\it et al.} (1993).
The parameters used in the turbulence model of these papers
were estimated from the laboratory experiments 
without the density stratification and the effects of the Coriolis 
and the tidal forces.

Sekiya (1998) obtained 
the quasi-equilibrium distributions of dust density, 
including the effect of the density stratification,
but neglecting the effects of the Coriolis and the tidal forces.  
The vertical shear rate $\partial v/\partial z$ increases 
as dust aggregates settle toward the midplane, where 
$z$ is the distance from the midplane.  
Eventually, 
the shear rate exceeds the critical value of the shear instability,
{\it i.e.,} the Richardson number becomes less than 0.25.
The Richardson number is defined by 
\begin{equation}
	J= - \frac{ g_z \partial \rho / \partial z}
	{\rho (\partial \bar{v} / \partial z)^2},
	\label{eq:ri}
\end{equation}
where $g_z$ is the vertical gravitational acceleration.
Then turbulence may be developed.  
If dust aggregates are small enough, very weak turbulence 
can stir up the dust aggregates and prevent them from settling further.  
Thus, the dust density distribution regulates itself 
so that the shear rate is marginally above the critical value $J_c =0.25$.  
Sekiya (1998) calculated analytically this density distribution 
with the critical shear rate.
He has concluded that the dust density hardly exceeds 
the critical density of the gravitational instability, 
as long as the dust to gas mass ratio inferred from the solar abundance 
is assumed.
He has shown that if the dust to gas mass ratio increases 
one order of magnitude by gas escape or dust concentration, 
the dust density at the midplane exceeds the critical value of the shear 
instability, and planetesimal can be formed 
through  gravitational fragmentation of the dust layer.  

Youdin and Shu (2002) also found the critical ratio of
 the surface density ratio of dust and gas
using the density distribution
 with the critical shear rate. 
In addition, they suggested a dust concentration mechanism 
through radial drift of chondrule-sized dust by gas drag in the nebula. 
This mechanism enables the density ratio of dust and gas 
to exceed the critical value in $\leq$ few $\times 10^6$ year.

The derivation of dust density distribution by Sekiya (1998),
and Youdin and Shu (2002) was somewhat intuitive.  
More elaborate calculations of the shear instability are needed 
in order to conclude that 
the gravitational instability model of planetesimal formation 
is denied.  
For the first step, Sekiya and Ishitsu (2000, 2001) (Hereafter 
Papers I and II) investigated  
the shear instability of the dust layer using the linear analysis, 
neglecting the effects of the Coriolis and the tidal forces.  

In Paper I,
we assumed that the unperturbed 
density had the constant Richardson number density distributions
by Sekiya (1998). 
The results show:
(A) The flow is stable for the Richardson number  $J \simge 0.22$.  
(B) The growth time of the shear instability 
is much longer than the Kepler period, 
as long as the Richardson number $ J \simge 0.1$.  
On the other hand, the Coriolis and the tidal forces 
would affect the flow in time scale on the order of 
the Kepler period.  
Thus the neglect of these forces is not good 
for the constant Richardson number density distribution 
with $ J \simge 0.1$. 

In Paper II,
 the linear stability analysis  was
 performed using  the following density distribution
which will be called the {\it hybrid density distribution}
(an inner constant density distribution, plus outer
sinusoidal transition regions):
\begin{equation}
\rho_{d0}(z) =
\left\{ \begin{array}{rl}
\rho_{d0}(0) & \mbox{for $|z|\le z_d-2h_d$,} \\
\rho_{d0}(0)\{1-\sin [\pi (z-z_d+h_d)/2h_d]\}/2 & 
\mbox{for $z_d-2h_d<|z|<z_d$,} \\
0 & \mbox{for $z_d\le |z|$},
\end{array}
\right.
\label{eq:rho0}
% Eq.(2)
\end{equation}
where $z_d$ the half-thickness of the dust layer,
and $h_d$  the half-thickness of the transition zones,
where the dust density varies from $\rho_{d0}(0)$ to 0 sinusoidally.
Here the half-thickness of the dust layer  is given by
\begin{equation}
	z_d =\frac{\Sigma_d  }{2\rho_{d0}(0)} + h_d,
\end{equation}
and the surface density of the dust is given by
\begin{equation}
\Sigma_d=\int_{-\infty}^{+\infty}\rho_d dz=
\left\{ \begin{array}{rl}
7.1f_d(r/\mbox{AU})^{-1.5}\mbox{g cm}^2\mbox{ for } r<2.8\mbox{AU},\\
30f_d(r/\mbox{AU})^{-1.5}\mbox{g cm}^2\mbox{ for } r>2.8\mbox{AU},
	\label{eqn:Sigma}
\end{array}
\right.
\end{equation}
where $\rho_d$ is the dust density defined by the total dust mass 
floating in a unit volume, and $f_d$ is a parameter 
($f_d=1$ for the Hayashi model). 
We used Hayashi's solar nebula model (Hayashi 1981, Hayashi 
{\it et al.}, 1985) at 1AU as the dust surface density $\Sigma_d$
in the most of calculations except for Figs. \ref{fig:fd},
\ref{fig:tfig15}, and \ref{fig:tfig16}.

In the following, we explain the reason why the hybrid density 
distribution is used.
Dust particles which are distributed uniformly at first stick
together to form dust aggregates. In a laminar nebula, the settling 
velocity $v_{dz}$ of a dust aggregate is given by
\begin{equation}
	v_{dz} = - \tau_{f} \Omega_K^2 z,
\end{equation}
where $\tau_{f}$ is the frictional time of the dust aggregate. 
Dust aggregates grow faster in regions with larger $|z|$,
since the principal relative velocity of dust aggregates
is induced by difference of settling velocities
of dust aggregates with different frictional times
(Weidenschilling 1980, Nakagawa {\it et al.} 1981).
As dust aggregates grow,
their settling velocities increase if the dust aggregates are compact. 
Thus, dust aggregates accumulate in a certain region
with an intermediate value of $|z|$
(see 1000 yrs and 1300 yrs density distribution
in Fig. 2 of Nakagawa {\it et al.} (1981)).
This state is unstable for the Rayleigh-Taylor instability,
and the dust density distribution is
considered to be adjusted as to be constant
in the dust layer (Watanabe and Yamada 2000).
Thus, the constant distribution in the dust layer as given 
by Eq. (\ref{eq:rho0}) for $|z|\le z_d-2h_d$ 
is considered to be a natural consequence 
of the dust settling in a laminar disk.
For simplicity, the sinusoidal density distribution 
is assumed in the transition region ($z_d-2h_d<|z|<z_d$).

According to results of Paper II, if $\rho_d(0)/\rho_g \sim 1$, 
the growth rate of the instability is 
on the order of the Keplerian angular frequency.  
On the other hand, 
if $\rho_d(0)/\rho_g >> 1$, the growth rate is 
much larger than the Keplerian angular frequency.  
Thus, we have expected  that the Coriolis and tidal forces 
might not have an important effect as long as 
$\rho_d(0)/\rho_g >> 1$.

Ishitsu and Sekiya (2002) (Hereafter Paper III),
 carried out calculations taking 
 the Coriolis force into account, but neglecting the tidal force. 
The results showed that the Coriolis force had little effects.
However, we found a new type of resonances  which resembles the 
Lindbland resonances if the growth rate is similar to
or smaller than the Keplerian angular frequency.
The energy source of  the instability is this resonance and different from
that of the shear instability.
These results will be reviewed in Section 4.

In this paper, we  solve linearized perturbation equations
with the tidal and the Coriolis forces under the following assumptions:
(1) The self-gravity is neglected.
(2) A mixture of gas and dust is treated as one fluid,
which is a good approximation in the case
where dust aggregate sizes are small ($\simle$ 1cm).
(3) The gas is incompressible
 since the dust layer 
which is much less than vertical scale height of the gas disk $H$ is treated.
(4) The effects of the radial density and the pressure gradient
of the unperturbed state are only incorporated
in the unperturbed rotation velocity distribution $v_0(z)$.
(5) Local Cartesian coordinates ($x, y, z$)  are used and we
neglect the curvature of a circle 
with constant values of $r$ and $z$.
(6) The hybrid dust density distribution used in Papers II and III
is  adopted as the unperturbed dust density distribution. 

When the dust layer with the hybrid dust density distribution 
becomes unstable for a linear perturbation, the linear mode grows. 
Eventually, nonlinear effects become important, 
and a quasi-equilibrium state between settling and
 turbulent diffusion of dust aggregates would be realized. 
This paper treats the first linear growth of the instability. 
Of course, it is very important to obtain quasi-equilibrium state 
with taking account of the Coriolis and the tidal forces, 
which is left in future works. Note that $J=$const distribution 
obtained Sekiya (1998) is not necessarily realize,
since it neglects the effects of the Coriolis and the tidal forces.

In Section II,  the basic equations for the linear analysis are derived. 
In Section III, a numerical method is described.
In Section IV, solutions with the Coriolis force but 
without the tidal force (Paper III) are reviewed.
In Section V, solutions with both the Coriolis and the tidal
forces  are given.
In Section VI, conclusions obtained in this paper are summarized.

\section{FORMULATION}
 We use the local Cartesian coordinate system around a radius $R$ 
 from the central star rotating with the 
Kepler angular frequency $\Omega_K(R)$:
 \begin{equation}
   x = r - R,
	\label{eqn:localx}
 \end{equation}
 \begin{equation}
   y = R[\phi -\Omega_K(R)t],
 \end{equation}
 \begin{equation}
   z = z,
 \end{equation}
 where the  curvature of $r=$ constant circle is neglected 
 since $O(\partial /\partial r) >> O(1/r)$ and 
 $( \partial /\partial{\phi}) /r \approx (\partial /\partial y)
$ for $ r>> |x|,|y|$.

 In the rotational system that revolves around a central star with 
  angular frequency $\Omega_K(R)$, the Coriolis and the centrifugal  
forces emerge. The tidal force, resultant force 
 of  gravity of central star and centrifugal force,
 is given in the local approximation regime by $3x \Omega_K^2(R)$.
  Hereafter, $\Omega_K(R)$ is denoted by $\Omega_K$ for simplicity.
Note that $\Omega_K$ is constant.
The hydrodynamic equations are given 
for the local Cartesian co-ordinates which move azimuthally with the Kepler 
velocity $v_K$ and rotate with the Kepler angular velocity $\Omega_K$ by
 \begin{equation}
	  \DP{u}{x} + \DP{v}{y} + \DP{w}{z} =0,
	 \label{eqn:ba1}
 \end{equation}
 \begin{equation}
	 \DP{\rho}{t} +u \DP{\rho}{x} + 
	 v\DP{\rho}{y} + w\DP{\rho}{z} =0,
	 \label{eqn:ba2}
 \end{equation}
\begin{equation}
	 \DP{u}{t} + u \DP{u}{x} + v\DP{u}{y} + w\DP{u}{z} 
      = - \frac{1}{\rho} \DP{P}{x} + 3 T \Omega_{K}^2 x +2 \Omega_{K}v.
	\label{eqn:moba3}
\end{equation} 
 \begin{equation}
	  \DP{v}{t} + u \DP{v}{x} + v \DP{v}{y} + w\DP{v}{z} 
       = - \frac{1}{\rho} \DP{P}{y}  -2 \Omega_{K} u,
	 \label{eqn:ba4}
 \end{equation}
 \begin{equation}
	  \DP{w}{t} + u \DP{w}{x} + 
	 v \DP{w}{y} +w \DP{w}{z} 
       = - \frac{1}{\rho} \DP{P}{z} -  \Omega_K ^2 z,
	 \label{eqn:ba5}
 \end{equation}
Note that only the case $T=1$ is realistic; on the other hand, the case $T=0$
corresponds to the model used in Paper III, where the tidal force is 
neglected. In this paper, calculations are performed for $0 \le T \le 1$,
in order to elucidate the effect of the tidal force on the instability 
in the dust layer. If $u=P=0$, Eq. (\ref {eqn:moba3}) reads 
\begin{equation}
v= - \frac{3}{2} T \Omega_K  x.
\end{equation}
This expresses the circular Kepler motion in the local coordinate system
(for $T=1$).
 In order to eliminate the Keplerian part of the velocity,
we introduce the velocity relative to the Keplerian motion.
\begin{equation}
\bar{v} = v + \frac{3}{2} T\Omega_K x.
\end{equation} 
From Eqs. (\ref{eqn:ba1}) to  (\ref{eqn:ba5}), we have
\begin{equation}
	 \DP{u}{x} + \DP{\bar{v}}{y} + \DP{w}{z} =0,
	\label{eqn:bta1}
\end{equation}
\begin{equation}
	\DP{\rho}{t} +u \DP{\rho}{x} + 
	(\bar{v} - \frac{3}{2} T\Omega_K x)\DP{\rho}{y} + w\DP{\rho}{z} =0,
	\label{eqn:bta2}
\end{equation}
\begin{equation}
	 \DP{u}{t} + u \DP{u}{x} + 
   (\bar{v} - \frac{3}{2} T\Omega_K x) \DP{u}{y} + w\DP{u}{z} 
      = - \frac{1}{\rho} \DP{P}{x} + 2 \Omega_{K}\bar{v},
	\label{eqn:bta3}
\end{equation}
\begin{equation}
	 \DP{\bar{v}}{t} + u \DP{\bar{v}}{x} + (\bar{v} - \frac{3}{2} 
        T\Omega_K x)\DP{\bar{v}}{y} + w\DP{\bar{v}}{z} 
      = - \frac{1}{\rho} \DP{P}{y}  - (2 - \frac{3}{2} T) \Omega_{K} u,
	\label{eqn:bta4}
\end{equation}
\begin{equation}
	 \DP{w}{t} + u \DP{w}{x} + 
 	(\bar{v} - \frac{3}{2} T\Omega_K x)  \DP{w}{y} +w \DP{w}{z} 
      = - \frac{1}{\rho} \DP{P}{z} -  \Omega_K ^2 z.
	\label{eqn:bta5}
\end{equation}

In order to carry out linear calculations, we assume 
that an unperturbed state is steady and uniform in $x$ and $y$ directions:
\begin{equation}
 \DP{}{t}= \DP{}{x}= \DP{}{y}= 0.
	\label{eqn:ntp.1}
\end{equation}
We also assume that the unperturbed velocity is 
parallel to $y$-axis (i.e. azimuthal direction)
\begin{equation}
 u_0 = w_0 = 0.
	\label{eqn:ntp.2}
\end{equation}
From Eqs. (\ref{eqn:bta3}),  (\ref{eqn:bta5}) and (\ref{eqn:ntp.2}), we have 
\begin{equation}
 \frac{1}{\rho_0}\DP{P_0}{x} = 2  \Omega_K {\bar{v}_0},
	\label{eqn:ntp.3}
\end{equation}
and
\begin{equation}
\frac{1}{\rho_0} \DP{P_0}{z} = - \Omega_K^2 z,
	\label{eqn:ntp.4}
\end{equation}
respectively. 

The unperturbed azimuthal velocity is calculated
from a given dust density distribution $\rho_{d0}$ and a given value 
of $\eta$  by using Eqs. (\ref{eqn:introv}) and (\ref{eqn:introrho}):
\begin{equation}
\bar{v}_0 = -\frac{\rho_g}{\rho_0} \eta v_K,
\label{eqn:np.5}
\end{equation}
where values of $\rho_g$, $\eta$, $v_K$  are assumed to be
constant, and $\rho_0$ depends only on $z$ and  is independent 
of $x$ in the local approximation with $r \approx R $.
The midplane gas density $\rho_g=\rho_g(0)$ is given by 
\begin{equation}
		\rho_g(0)=\sqrt{\pi} \Sigma_g /H, 
\end{equation}
where $\Sigma_g$ is surface gas density:
\begin{equation}
		\Sigma_g = 1700f_g(r/\mbox{AU})^{-1.5}\mbox{g cm}^2,
	\label{eqn:Sigmag}
\end{equation}
and $f_g$ is a parameter ($f_g=1$ for the Hayashi model).
The radial pressure gradient $\partial P_0 /\partial x$ is then
given by  Eq. (\ref{eqn:ntp.3}).
%$B@~7A2=(B

Linearizing Eqs. (\ref{eqn:bta1}) to (\ref{eqn:bta5})
and using Eqs. (\ref{eqn:ntp.3}) and (\ref{eqn:ntp.4}), we have
\begin{equation}
	 \DP{ u_1}{x} + \DP{v_1}{y} + \DP{w_1}{z} =0,
	\label{eqn:tpe.1}
\end{equation}
\begin{equation}
	\DP{ \rho_1}{t} + (\bar{v}_0-\frac{3}{2}T \Omega_Kx)\DP{\rho_1}{y}
+\DD{\rho_0} {z}w_1=0,
		\label{eqn:tpe.2}
\end{equation}
\begin{equation}
	  \DP{ u_1}{t}+(\bar{v}_0-\frac{3}{2}T \Omega_K x)\DP{u_1}{y}
      = - \frac{1 }{\rho_0} \DP{P_1}{x} + 2  \Omega_K \frac{\bar{v}
         _0}{\rho_0}\rho_1 + 2 \Omega_{K}v_1,
		\label{eqn:tpe.3}
\end{equation}
\begin{equation}
       \DP{ v_1}{t}+(\bar{v}_0-\frac{3}{2}T \Omega_K x)\DP{v_1}{y}
	+\DD{v_0}{z}w_1 
      = - \frac{1}{\rho_0} \DP{P_1}{y} - (2-\frac{3}{2}T) \Omega_{K}u_1 
		\label{eqn:tpe.4}
\end{equation}
\begin{equation}
 \DP{ w_1}{t}+(\bar{v}_0-\frac{3}{2}T \Omega_K x)\DP{w_1}{y}
  = - \frac{ 1}{\rho_0} \DP{P_1}{z}
  - \frac{\Omega_K ^2  z}{\rho_0} \rho_1.
		\label{eqn:tpe.5}
\end{equation}

 The normal mode analysis 
({\it i.e.} the Fourier transform) for $x$ cannot be done in this coordinates 
system because the coefficients  of equations depend on $x$.
Thus we transform coordinate $y$ into the shearing coordinate $y'$ 
with velocity $-\frac{3}{2} T \Omega_K x$ in order that the normal mode 
analysis for $x$ can be done (see e.g., Goldreich and Lynden-Bell 1965,
Goldreich and Tremaine 1978,  Ryu and Goodman 1992):
\begin{equation}
   y'= y + \frac{3}{2}T \Omega_K x t.
\end{equation}
We assume that  perturbed quantities are written
\begin{equation}
   f_1(x,y,z,t) = \hat{f}_1(z,t) \exp[i (k_y  y'+ k_x x)].
	\label{eqn:exp}
\end{equation}
Substituting this expression into Eqs. (\ref{eqn:tpe.1})--(\ref{eqn:tpe.5}),
the perturbation equations are rewritten
\begin{equation}
	 ik'_xu_1 + i k_y v_1 + \DP{w_1}{z} =0,
	\label{eqn:pt1}
\end{equation}
\begin{equation}
	\DP{ \rho_1}{t} +  i k_y \bar{v}_0 \rho_1
+\DD{\rho_0} {z}w_1=0,
	\label{eqn:pt2}
\end{equation}
\begin{equation}
	  \DP{ u_1}{t}+ i k_y \bar{v}_0  u_1
      = - i k'_x \frac{1}{\rho_0} P_1 + 2\Omega_K 
       \frac{\bar{v}_0}{\rho_0}\rho_1 + 2 \Omega_{K}v_1,
	\label{eqn:pt3}
\end{equation}
\begin{equation}
       \DP{ v_1}{t}+ i k_y \bar{v}_0  v_1
      = - i k_y \frac{1}{\rho_0} P_1  - \DD{\bar{v}_0}{z} w_1
      - \left(2 - \frac{3}{2}T\right) \Omega_{K} u_1 
	\label{eqn:pt4}
\end{equation}
\begin{equation}
 \DP{ w_1}{t}+ i k_y \bar{v}_0  w_1
  = - \frac{ 1}{\rho_0} \DP{P_1}{z}
  - \frac{\Omega_K ^2  z}{\rho_0} \rho_1,
	\label{eqn:pt5}
\end{equation}
where,
\begin{equation}
	 k'_x = k_x +  \frac{3}{2}T k_y \Omega_K t.
	\label{eqn:pt6}
\end{equation}
Here, we cannot use the Fourier transform method with respect to t  
 because the coefficients of these equations depend on $t$.
The linearized equations must be integrated numerically by using the finite
difference method  in order to see the growth  of an instability.

% Boundary conditions

Boundary conditions for $z$-direction are given as follows.
Only odd solutions for $w_1$ ({\it i.e.} $w_1=0$ at $z=0$) are considered 
since even ones  are always stable according to  our calculations.
The continuity of pressure at the boundary between the dust and gas layers
were applied in the analytical method of Papers I to III,
but it is difficult to apply this condition to our numerical method.
Thus, we solve the perturbation equation numerically within
 region $[0,z_0]$ where the solid-wall condition is applied  
at the boundary for simplicity. The value of $z_0$ should be large enough
 in order for an eigenfunction to decay sufficiently at a boundary $z_0$. 
It was confirmed that numerical solutions for $T=0$ with this condition 
 agree well with  solutions of the Paper III.
Thus, the  boundary conditions  are given by
\begin{equation}
 w_1 = 0  \; \mbox{ at }  \;  z= 0 \mbox{ and } z_0.
	\label{eqn:bt1}
\end{equation}
From Eqs. (\ref{eqn:pt5}) and (\ref{eqn:bt1}), we have
\begin{equation}
 \DP{P_1}{z} + \Omega_K^2 z \rho_1 = 0  \; \mbox{ at }  \;  
z=0 \mbox{ and } z_0.
\end{equation}
From Eqs. (\ref{eqn:pt2}) and (\ref{eqn:bt1}), we have
\begin{equation}
 \DP{\rho_1}{t} + i k_y \bar{v}_0 \rho_1= 0 
 \; \mbox{ at }  \;  z= 0 \mbox{ and } z_0.
	\label{eqn:brho}
\end{equation}
If we give an initial condition with  $\rho_1=0$ at $z=0 \mbox{ and } z_0$,
Eq. (\ref{eqn:brho}) implies that $\rho_1=0$ for $ t \geq 0 $.
Hereafter we take only such initial conditions for simplicity.
 Accordingly boundary conditions for $P_1$ and $\rho_1$ are given by
\begin{equation}
 \DP{P_1}{z} = 0  \; \mbox{ at }  \;  z= 0 \mbox{ and } z_0,
\end{equation}
\begin{equation}
\rho_1= 0  \; \mbox{ at }  \;  z= 0 \mbox{ and } z_0.
\end{equation}

\section{NUMERICAL METHOD }
\hspace*{\parindent}
We adopt MAC method (Harlow and Welch 1965), {\it i.e.}
pressure is determined by the condition that the equation of continuity
is satisfied in the next step, and other variables $u_1,v_1,
w_1$ and $\rho_1$ are calculated using this value of  pressure. 

We define  the divergence of a perturbed velocity by
\begin{equation}
	D \equiv ik'_x u_1 + i k_y v_1 + \DP{w_1}{z}.
	\label{eqn:div}
\end{equation}
Multiplying Eq. (\ref{eqn:pt3}) by $ i k'_x $ and  Eq.
(\ref{eqn:pt4}) by $i k_y$,  and  taking partial derivative 
Eq. (\ref{eqn:pt5}) with respect to $z$, and adding, we have
\begin{eqnarray}
	\DP{D}{t} &=& 
- i k_y \bar{v}_0 D - 2 i k_y \DD{\bar{v}_0}{z} w_1  
-\frac{1}{\rho_0} \left( -k'^2_x -k^2_y  
+  \DP[2]{}{z} \right)P_1 + \frac{1}{\rho_0^2} \DD{\rho_0}{z} 
 \DP{P_1}{z} \nonumber \\
&& + 2 i  \Omega_K k'_x \frac{\bar{v}_0} {\rho_0} \rho_1
+ 2 i \Omega_K  k'_x  v_1  -i \left( 2 - 3T \right)\Omega_K k_y  u_1  
+ \Omega_K^2 \DD{}{z} \left( - \frac{z}{\rho_0} \rho_1 \right).
\label{eqn:div2}
\end{eqnarray}
We use the first order approximation:
\begin{equation}	
	\DP{D}{t} \approx \frac{D^{n+1} -D^{n}}{\Delta t}.
	\label{eqn:eul}
\end{equation}
We require that the equation of continuity (\ref{eqn:pt1}) is satisfied
in the next step, {\it i.e.} $D^{n+1}=0$. Thus, we get  the  Poisson equation
 for $P_1$:
\begin{eqnarray}
&& \left( -k'^{n2}_x -k^2_y  +  \DP[2]{}{z}  -
\frac{1}{\rho_0} \DD{\rho_0}{z}  \DP{}{z}   \right)P_1^n \nonumber \\	
&=& \rho_0 \left\{ \left(\frac{1}{\Delta t} - 2 i k_y \bar{v}_0 \right) D^n
 - 2 i k_y \DD{\bar{v}_0}{z} w_1^n   + 2 i \Omega_K  k'^n_x  v_1^n
  -i \left(2 -3T\right)\Omega_K  k_y  u_1^n \right.
\nonumber \\	
&& \left. + 2 i \Omega_K   k'^n_x \frac{\bar{v}_0}{\rho_0} \rho_1^n 
+ \Omega_K^2 \DD{}{z}\left( - \frac{z}{\rho_0} 
\rho_1^n \right) \right\}.
\label{eqn:poisson}
\end{eqnarray}
From this equation, we can get $P_1$ at the time step $n$. Next, we calculate
$\rho_1,u_1,v_1$ and $ w_1$ at the time step $n+1$ from 
Eqs. (\ref{eqn:pt2})--(\ref{eqn:pt5}): 
\begin{equation}
	 \rho_1^{n+1} =  \rho_1^{n}  
+ \Delta t \left\{-  i k_y \bar{v}_0 \rho_1^n
 - \DD{\rho_0} {z}w_1^{n} \right\},
	\label{eqn:spt2}
\end{equation}
\begin{equation}
	   u_1^{n+1} =  u_1^{n}  + \Delta t \left\{ -  i k_y \bar{v}_0  u_1^n
       - i k'^n_x \frac{1 }{\rho_0} P_1^n
    + 2 \Omega_K \frac{\bar{v}_0}{\rho_0}\rho_1^n
	+ 2 \Omega_{K}v_1^n \right\},
	\label{eqn:spt3}
\end{equation}
\begin{equation}
       v_1^{n+1} =  v_1^{n}  + \Delta t \left\{ -i k_y \bar{v}_0 v_1^n
	- \DD{\bar{v}_0}{z} w_1^n
       - i k_y \frac{1}{\rho_0} P_1^n
      - \left(2 - \frac{3}{2}T\right) \Omega_{K}u_1^n  \right\},
	\label{eqn:spt4}
\end{equation}
\begin{equation}
 w_1^{n+1} =  w_1^{n}  + \Delta t \left\{ - i k_y \bar{v}_0  w_1^n
   - \frac{ 1}{\rho_0} \DP{P_1^n}{z}
  - \frac{\Omega_K ^2  z}{\rho_0} \rho_1^n \right\}.
	\label{eqn:spt5}
\end{equation}
The above method is  the first order accuracy with respect to $\Delta t$.
In order to improve them to the second order accuracy, we use
the following strategy. 
 We replace   perturbed quantities $\mbox{\boldmath{$f$}}_1^n
\equiv (\rho_1^n, u_1^n, v_1^n, w_1^n)$ on right hand  side
of Eq. (\ref{eqn:poisson}) with 
\begin{equation}
 \mbox{\boldmath{$f$}}_1^{n+ \frac{1}{2} }   
= (\mbox{\boldmath{$f$}}_1^{n+1} + \mbox{\boldmath{$f$}}_1^{n}) /2.
\end{equation}
Again, we solve Eq. (\ref{eqn:poisson}) replacing $n$ with $n+ 1/2$.
More exact values of $\rho_1^{n+1},u_1^{n+1},v_1^{n+1}$ and $w_1^{n+1}$ are
obtained by replacing these  quantities at $n$ in braces of the right 
hands of Eqs. (\ref{eqn:spt2}) to (\ref{eqn:spt5})  
with ones at $n+1/2$.
We adopt one dimension staggered grid, where
 $w_1$ are estimated at grids and $P_1,\rho_1,u_1,v_1$  at midpoints 
of adjacent  grids.  
In Eqs. (\ref{eqn:poisson}), (\ref{eqn:spt2}) and (\ref{eqn:spt4}),
$w_1$ is calculated by taking the mean values at the adjent meshes.
So as $\rho_1$ in Eq. (\ref{eqn:spt5})

Numerical parameters ($\Delta t \Omega_K$, $z_0/\eta r$, Grid number 
for $z$-direction $N_z$)  
and model parameters ($\rho_d(0)/ \rho_g$, $h_d/z_d$, $f_g$,
 $f_d$, $k_x\eta r$, $\log(k_y^2\eta^2 r^2)$, $T$)
in this work are listed in Table I.
The minimum values of the Richardson number $J_{min}$ for a region $[0,z_0]$
are also listed as an indicator of shear strength.

Initial conditions were set as follows.
 Some  Fourier components of lower orders were selected 
as to satisfy boundary conditions of $u_1, w_1$ and $\rho_1$.
Each Fourier coefficient is given by a random number.
Velocity $v_1$ was determined from  $u_1$ and $w_1$ by using the 
equation of  continuity (\ref{eqn:pt1}).

\section{SHEAR INSTABILITY WITHOUT THE TIDAL FORCE}

Here, solutions without the tidal force, that is, the case  $T=0$ 
 is obtained in order to compare the results 
with the  tidal force (detailed description is given in Paper III).
Then, we can carry out the normal mode analysis for time:
\begin{equation}	
	\hat{f}_1(z,t) = \tilde{f}_1(z) \exp(-i \omega t),
\end{equation}
where $\omega$ is the angular frequency of perturbed quantities,
and obtain eigenvalues $\omega$ by solving the differential 
equations to satisfy with  boundary conditions. 
The growth rate of the instability is $\omega_I = \Im(\omega)$,
where $\Im$ denotes the imaginary part.

\begin{figure}[p]
\centerline{ 
\epsfbox{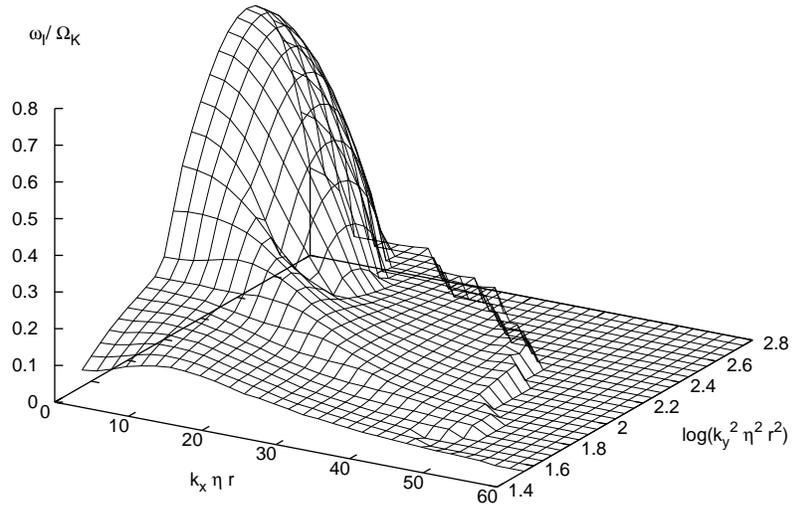}
}
\caption{The growth rate $\omega_I$ of the mode 
as a function of radial and azimuthal wave number $k_x$ and $k_y$
in the case where $T=0$ with $\rho_{d0}(0)/\rho_g=0.5$ and $h_d/z_d=0.2$.}
\label{fig:dcfig1}
\end{figure}

Figure \ref{fig:dcfig1}  shows the growth rate $\omega_I$ 
as a function of radial and azimuthal wave number $k_x$ and $k_y$
in the case where $\rho_{d0}(0) /\rho_g= 0.5$, and $h_d/z_d=0.2$.
The growth rate $\omega_I$  has a finite positive value 
for $ k_x <  k_{xc} $,
and $\omega_I=0$ for $ k_x >  k_{xc} $, where $k_{xc}$ is the critical 
radial wave number.
This is very important for the stability with  the tidal force
as will be written in the next section.

Figure \ref{fig:dcfig2} shows the growth 
rate of the instability with the most unstable wave number
(hereafter called ``the peak growth rate'') as a function of 
$\rho_d(0)/\rho_g$ with  $h_d/z_d=0.2$.
Note the slope of curve in Fig. \ref {fig:dcfig2} changes 
at $\rho_d(0)/\rho_g=$ 0.23 and 0.39. 
This is because one mode which has the peak growth rate 
at smaller dust density at the midplane 
moves to another mode at the two dust densities at the midplane.
The increase of the dust density  corresponds to 
the dust settling to the midplane in a quiescent nebula.
Thus, in the case without the tidal force, the growth rate of the
shear instability increases as the settling proceeds.
The peak growth rate also increases
when the half-thickness of the transition zone of the dust layer $h_d$ 
decreases (Fig. \ref{fig:dcfig4}). 

Figure \ref{fig:fd} shows the peak growth rate  
as a function of $f_d$ (see Eq. (\ref{eqn:Sigma})).
As $f_d$ increases, the peak growth rate decreases.
When $f_d$ increases with a constant dust density at midplane, 
the half-thickness of  transitional zone of the dust layer increases.
This implies that the shear rate $dv_0/dz$ decreases.
Thus, the shear instability is depressed if the dust surface
density increases due to some mechanism.

Figure \ref{fig:fg} shows the peak growth rate 
as a function of $f_g$ (see Eq. (\ref{eqn:Sigmag})).
In contrast with  $f_d$, as $f_g$ decreases, the peak growth rate increases.
Figure \ref{fig:fd} is identical to Figure \ref{fig:fg} when
1/$f_g$ is taken as $f_d$.
Thus, only $f_d$ will be taken as a numerical  paramter in the following.

\section{SHEAR INSTABILITY WITH THE TIDAL FORCE}
We performed a test run for the case $T=0$ in order to check our code, since
we have already gotten eigenfuctions of growing modes
in the previous section.
Figure \ref{fig:tfig1} shows time evolutions of the radial, 
azimuthal and vertical  components of the mean kinetic energy for
$\log(k_y^2\eta^2 r^2)=2.46$, $T=0$, $\rho_{d0}(0) /\rho_g= 0.5$ 
and $h_d/z_d=0.2$.
Note that important values are not the absolute values of 
components of the perturbed kinetic energy but ratios and 
the growth rate of those because we treat linearized quantities.
If a linearized variable grows proportional to $\exp(\omega_I t)$,
its absolute value squared varies as $f_1f_1^* \propto \exp(2\omega_I t)$,
where the superscript ${}^*$ denotes the complex conjugate value.
Thus, we calculate the growth rate by
\begin{equation}
	\omega_I = \frac{1}{2\Delta t}
 	\ln\left[\frac{(<u_1 u_1^* > +<v_1 v_1^* > + <w_1 w_1^* >)
       |_{t+\Delta t}}
	{(<u_1 u_1^*> +<v_1 v_1^* > + <w_1 w_1^* >)|_{t}}\right].
\end{equation} 
The time evolution of the growth rate  is shown 
in Fig. \ref{fig:tfig2}.
We find that the growth rate asymptotically approaches the analytical 
solution (expressed by the dotted line).
In addition, we have checked that numerically calculated 
function $w_1(z,t)$ for large values of $t$
is  almost same as the eigenfunction.
Thus we consider our code works correctly.

Figure \ref{fig:tfig3} shows the time evolutions of   
radial, azimuthal and vertical  components of mean kinetic energy
with same parameters as Fig. \ref{fig:tfig1} except that $T=1$.
The solid curve in Fig. \ref{fig:tfig5} shows the 
temporal growth rate with $\log(k_y^2\eta^2 r^2)=2.46$. 
When $T=1$, the growth rate  rises  
at first and then begins to vibrate, being zero  in average.
Thus the tidal force has a stabilizing effect.

The dotted curve in Fig. \ref{fig:tfig5} shows  the temporal growth rates 
with $\log(k_y^2\eta^2 r^2)=2.00$.
The temporal growth rate is  rather  insensitive to the wave number. 
The solution with a higher wave number has a tendency to transit 
to simple vibration earlier.
Initial conditions have influence on  growth  of the energy 
at the beginning.
However, subsequently, time evolution of the  growth rate and energy become 
independent of  initial conditions.

Figure \ref{fig:tfig6} shows the growth rate with different values 
of the tidal parameter $T$. Note that $T= 1$ is 
only the realistic case, and we solved cases with other values of $T$
for comparison in order to elucidate the effect of the tidal force.
As $T$ increases, the both  amplitude  and period of oscillation of $\omega_I$
increase. We have checked that these oscillation frequencies are almost equal 
to the epicyclic frequency, {\it i.e.} the oscillation frequency 
due to the Coriolis and the tidal forces 
(see {\it e.g.} Eq. (8.9) of Shu 1992)
\begin{equation}
	\kappa = \sqrt{ 2 \left( 2 -\frac{3}{2}T \right)} \Omega_K,
\end{equation}
and are different from  the tidal shear rate $3T\Omega_K/2$.
Here we consider the reason why the instability is stabilized even if 
$T$ is as small as 0.1. As found from Eq. (\ref{eqn:pt6}), $k'_x $ 
increases as time $t$ elapses.
The analysis for $T=0$ indicates that, as $k_x$  increases, 
the growth rate of the instability decreases and eventually
the instability  disappears (see Fig. \ref{fig:dcfig1}).
Consequently for $T \neq 0$, even if an unstable mode had grown
 at the beginning, it would be stabilized  as time elapses.
A time  for  stabilization $t_s$ is  estimated as follows.
Given the critical wave member $k_{xc}$ above which  $\omega_I=0$ for the case 
of $T=0$, we have
\begin{equation}
	 t_s = (k_{xc} -k_x) /\left( \frac{3}{2} k_y \Omega_K T \right),
	\label{eqn:crit} 
\end{equation}
from Eq. (\ref{eqn:pt6}).
It is understood from Eq. (\ref{eqn:crit}) that  
the time scale to  stabilize the instability decreases 
as wave number $k_y$ increases.
The oscillations of kinetic energy 
 and the growth rate are  interpreted as mutual interference  
of non-growing waves with $\omega_I =0$. 
We have found that the oscillation period hardly change 
when $k_x > k_{xc}$ in the case of $T=0$.
Indeed, the oscillation period for 
 $\rho_{d0}(0)/\rho_g =0.5$, $h_d/z_d =0.2$,
 $T=0$ and $k_x \eta r=50$ becomes almost same as 
that for $T=1$ as time elapses 
(compare Figs. \ref{fig:tfig3} and \ref{fig:tfig11}), although 
 the oscillation for $T=1$ seems different from  that for $T=0$
at the early phase due to relicts of initial conditions.

\begin{figure}[p]
\centerline{ 
\epsfysize=12cm
\epsfbox{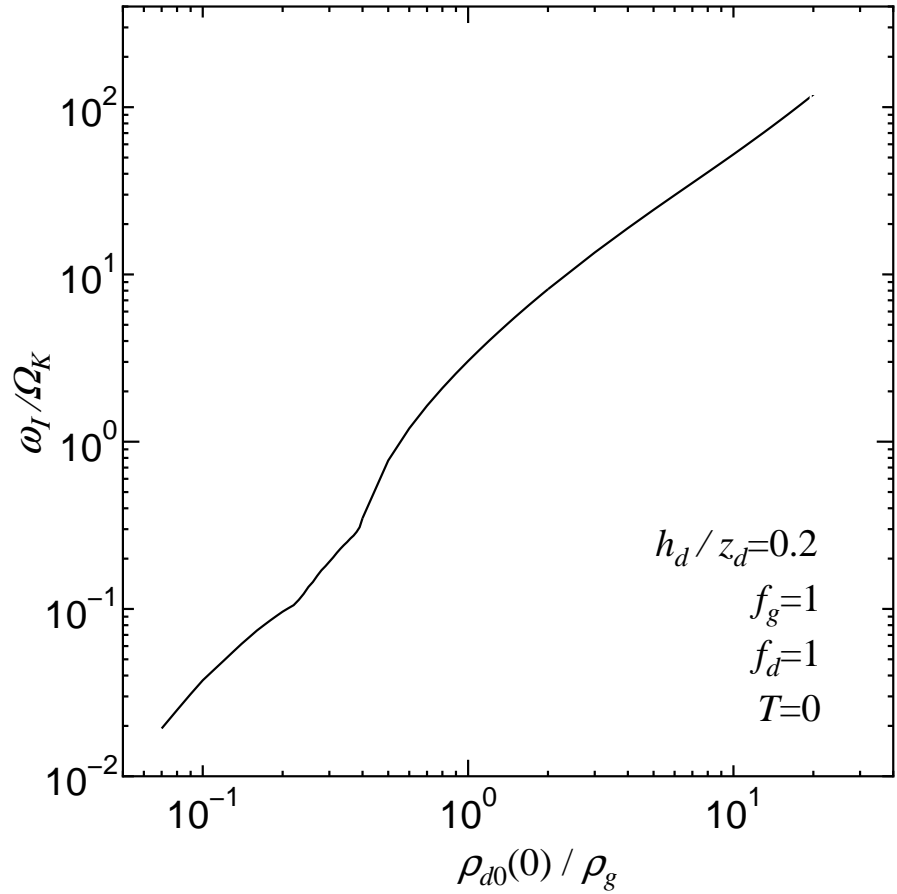}
}
\caption{The growth rate $\omega_I$ of the mode with the most unstable 
wave number as a function of the ratio of dust to gas on the midplane
$\rho_{d0}(0)/\rho_g$, in the case where $T=0$ with $h_d/z_d=0.2$. }
\label{fig:dcfig2}
\end{figure}

\begin{figure}[p]
\centerline{
\epsfysize=12cm
\epsfbox{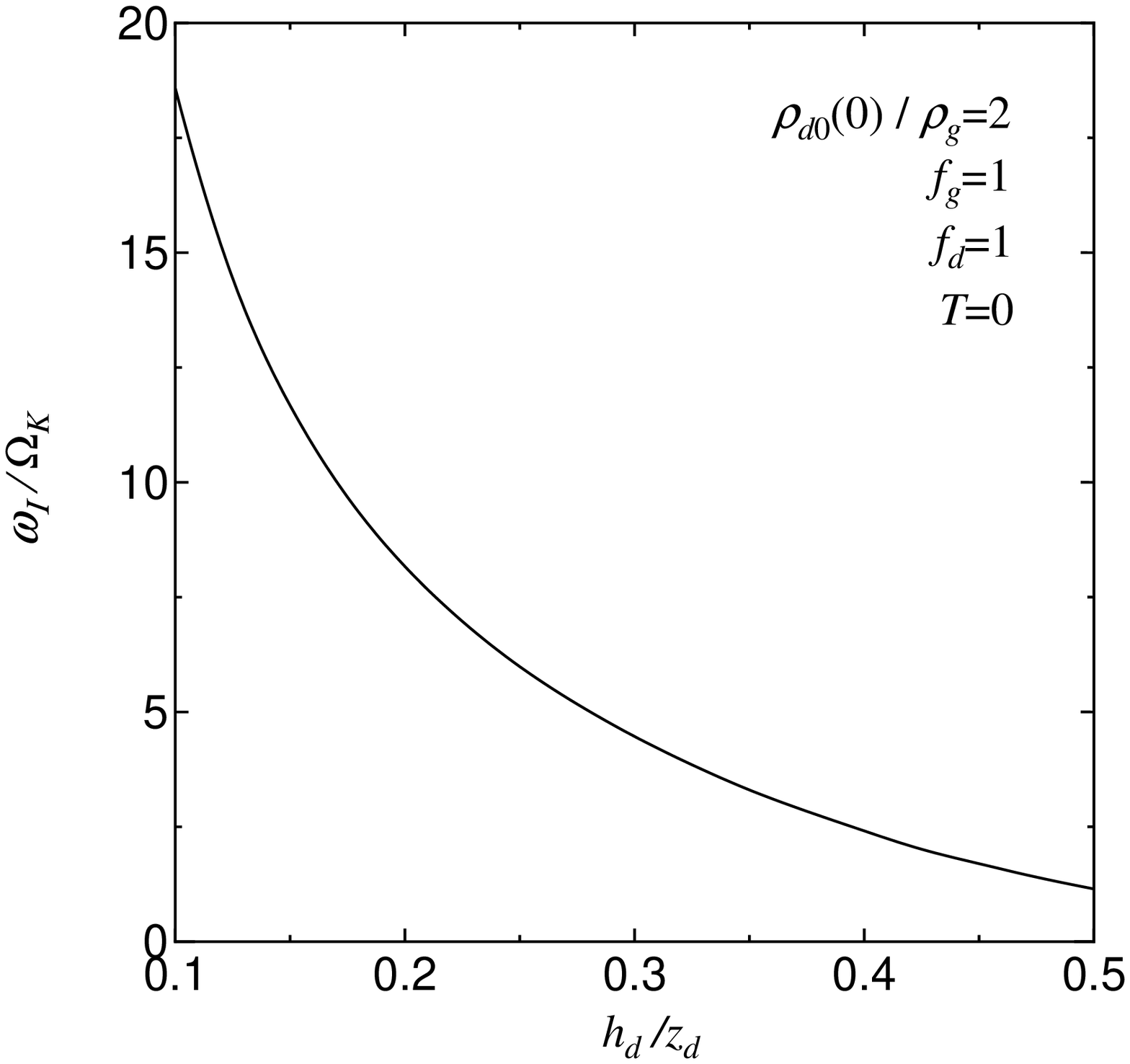}
}
\caption{The growth rate of instability $\omega_I$ of the mode with the most unstable 
wave number as 
a function  of the $h_d/z_d$
in the case where $T=0$ with $\rho_{d0}(0)/\rho_g=2$. }
\label{fig:dcfig4}
\end{figure}

\begin{figure}[p]
\centerline{
\epsfysize=12cm
\epsfbox{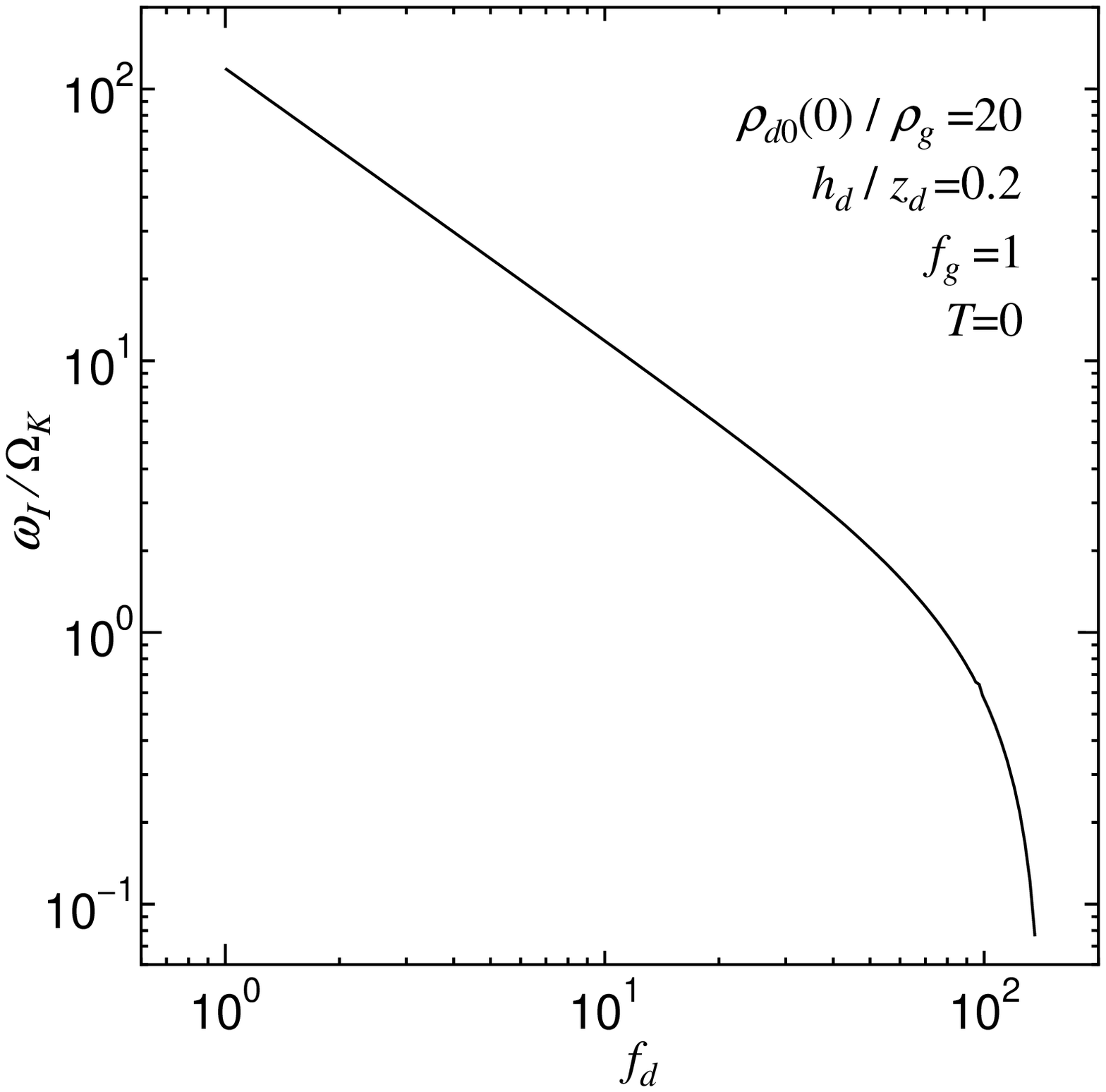}
}
\caption{The growth rate of instability $\omega_I$ of the mode with the most unstable 
wave number as 
a function  of the $f_d$
in the case where $T=0$ with $\rho_{d0}(0)/\rho_g=20$ and $h_d/z_d=0.2$. }
\label{fig:fd}
\end{figure}

\begin{figure}[p]
\centerline{
\epsfysize=12cm
\epsfbox{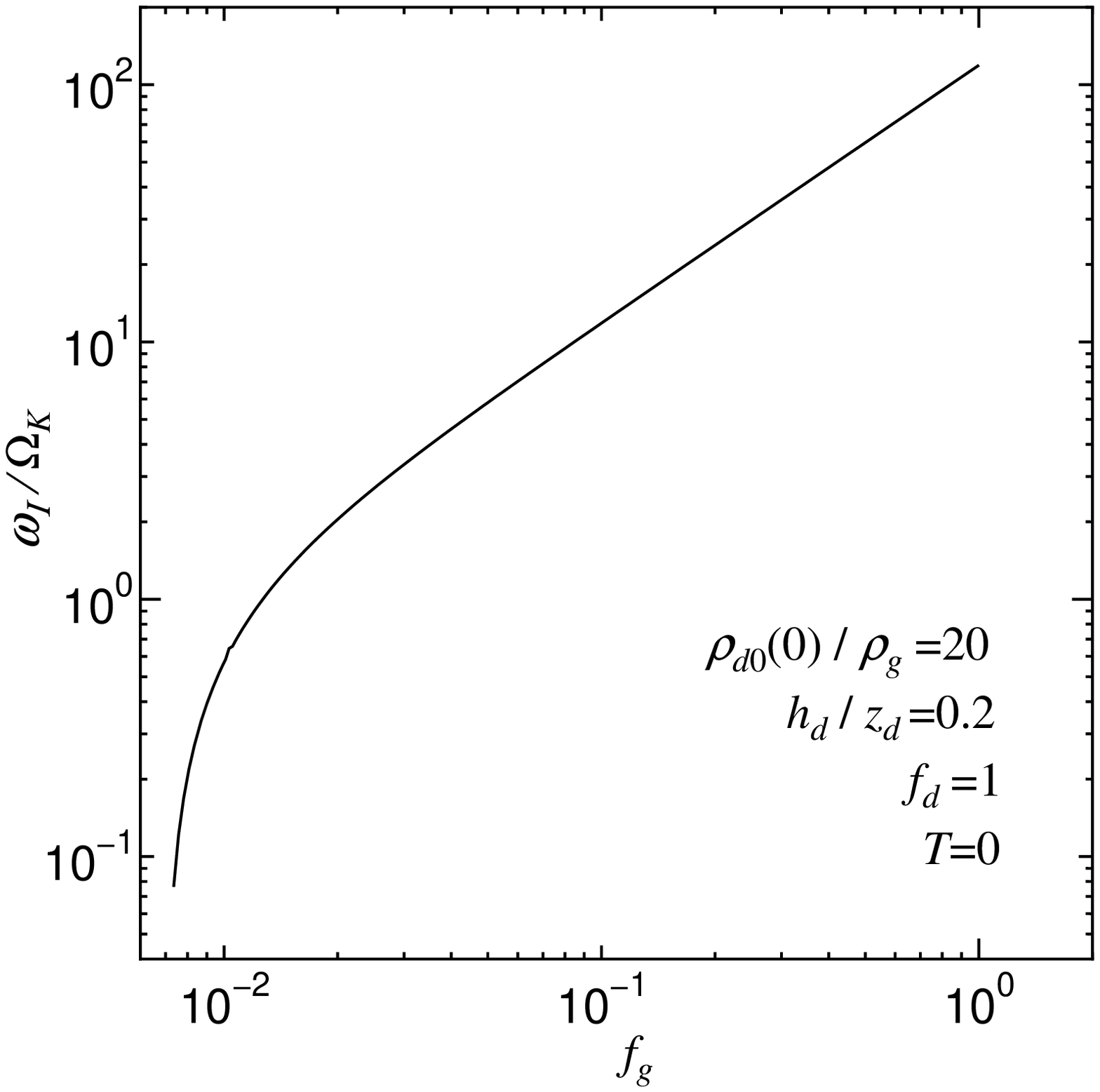}
}
\caption{ The growth rate of instability $\omega_I$ of the mode with the most unstable 
wave number as 
a function  of the $f_g$
in the case where $T=0$ with $\rho_{d0}(0)/\rho_g=20$ and $h_d/z_d=0.2$.}
\label{fig:fg}
\end{figure}

\begin{figure}[p]
\centerline{
\epsfysize=12cm
\epsfbox{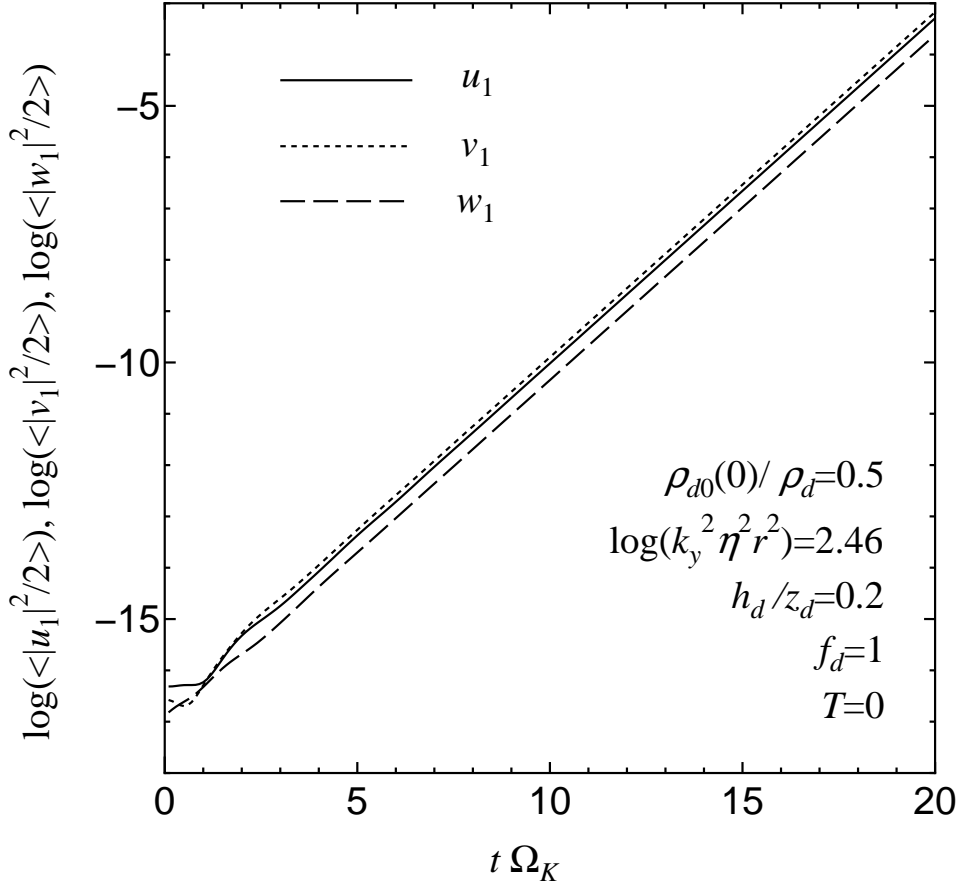}
}
\caption{The time evolution of  radial, azimuthal and vertical
parts of the perturbed kinetic energy density, in the case where
 $T=0$, with  $\log(k_y^2 \eta^2 r^2)=2.46$, 
$\rho_{d0}(0)/\rho_g=0.5$ and $h_d/z_d=0.2$.}
\label{fig:tfig1}
\end{figure}

\begin{figure}[p]
\centerline{
\epsfysize=12cm
\epsfbox{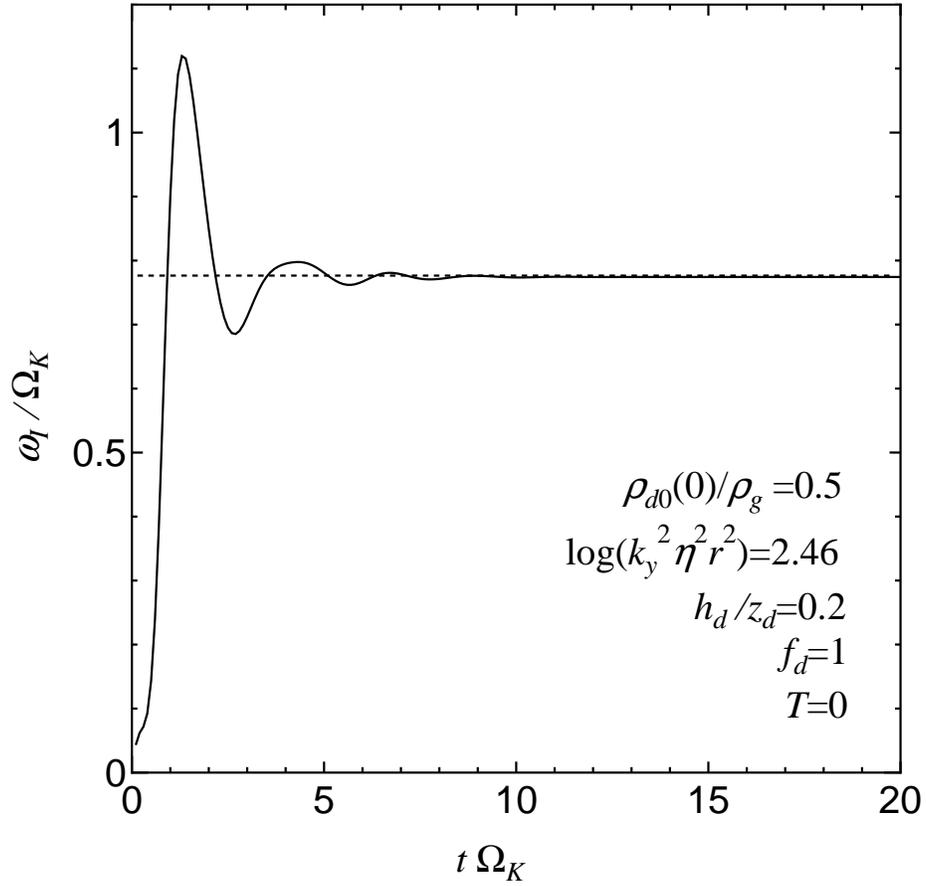}
}
\caption{The time evolution of  the growth rate, in the case where
 $T=0$, with  $\log(k_y^2 \eta^2 r^2)=2.46$, 
$\rho_{d0}(0)/\rho_g=0.5$ and $h_d/z_d=0.2$.
The dotted line shows the  analytical solution.}
\label{fig:tfig2}
\end{figure}

\begin{figure}[p]
\centerline{
\epsfysize=12cm
\epsfbox{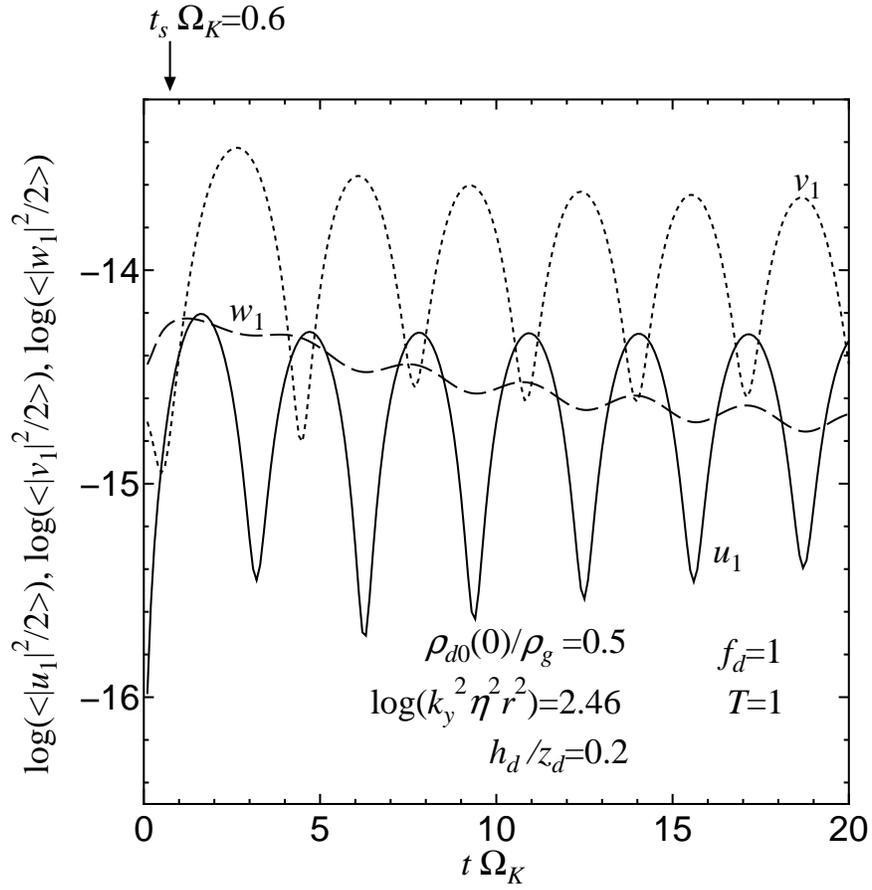}
}
\caption{Same as Fig.  \protect\ref{fig:tfig1} except that $T=1$
and the position of a arrow shows the stabilization time $t_s \Omega_K$.}
\label{fig:tfig3}
\end{figure}

\begin{figure}[p]
\centerline{
\epsfysize=12cm
\epsfbox{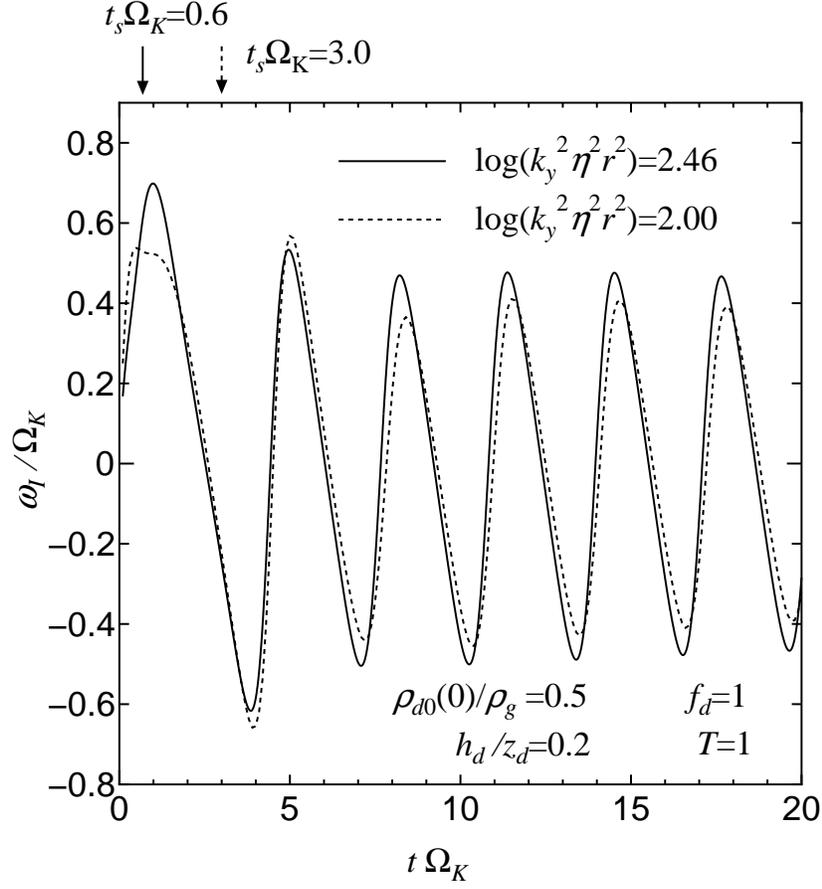}
}
\caption
{ The time evolution of  the growth rate for 
 $\log(k_y^2 \eta^2 r^2)=2.46$ ($i.e.$ the parameters have same values
as Fig. \protect\ref{fig:tfig3}) 
and $\log(k_y^2 \eta^2 r^2)=2.00$,
 with  $T=1$, $\rho_{d0}(0)/\rho_g=0.5$ and $h_d/z_d=0.2$.
The positions of solid and dotted arrows  show
 the stabilization times $t_s \Omega_K$ for $\log(k_y^2 \eta^2 r^2)=2.46$
and 2.00, respectively.}
\label{fig:tfig5}
\end{figure}

\begin{figure}[p]
\centerline{
\epsfysize=12cm
\epsfbox{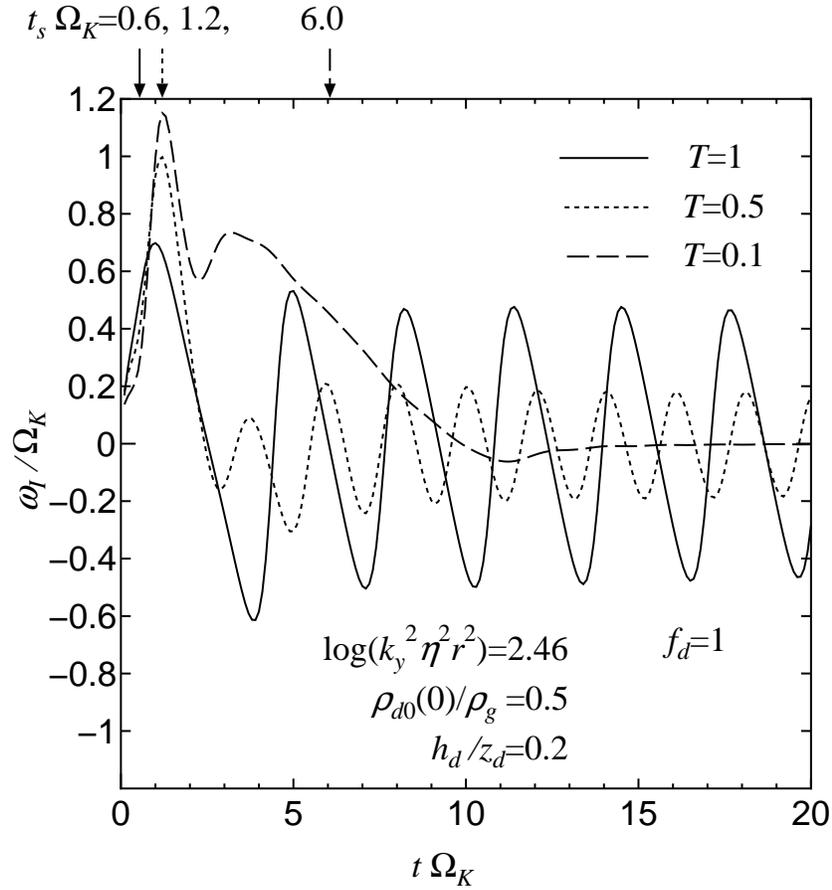}
}
\caption
{The time evolutions of  the growth rate for 
 $T=1$, 0.5 and 0.1, with  $\log(k_y^2 \eta^2 r^2)=2.46,
 \rho_{d0}(0)/\rho_g=0.5$ and $h_d/z_d=0.2$.
The positions of solid, dotted, and dashed arrows  show
 the stabilization times $t_s \Omega_K$ for $T= 1,$ 0.5 and 0.1, respectively.}

\label{fig:tfig6}
\end{figure}

\begin{figure}[p]
\centerline{
\epsfysize=12cm
\epsfbox{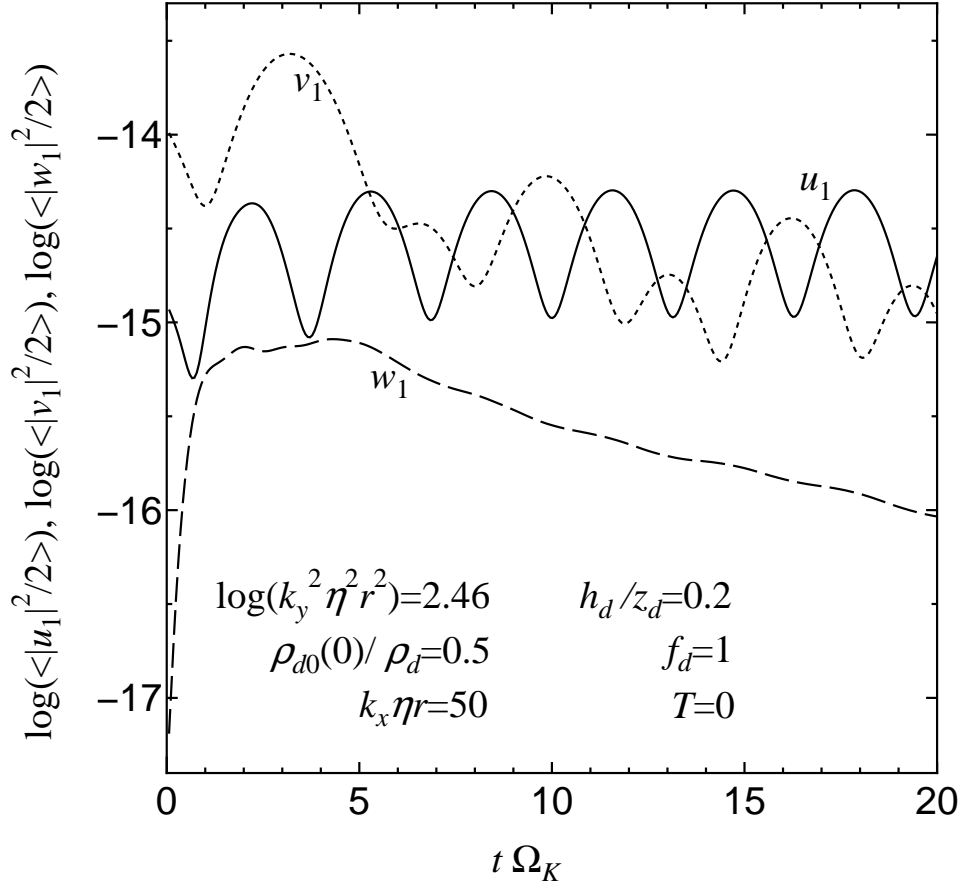}
}
\caption
{The time evolution of   radial, azimuthal and vertical
parts of the perturbed kinetic energy density
for  $T=0$ and $k_x \eta r = 50$, with  $\log(k_y^2 \eta^2 r^2)=2.46$,
 $\rho_{d0}(0)/\rho_g=0.5$ and $h_d/z_d=0.2$.}
\label{fig:tfig11}
\end{figure}

\begin{figure}[p]
\centerline{
\epsfysize=12cm
\epsfbox{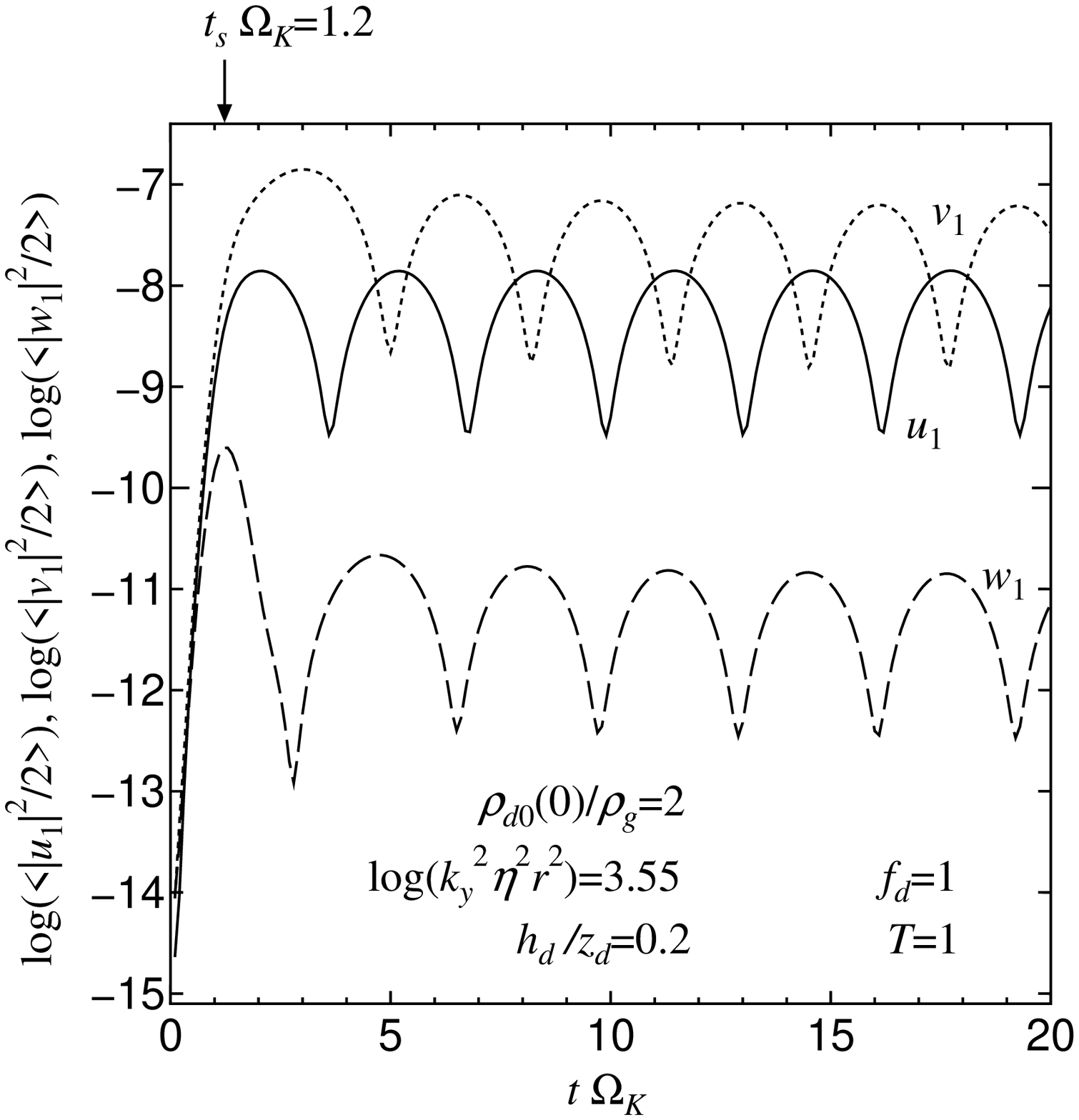}
}
\caption {The time evolution of  radial, azimuthal and vertical
parts of the perturbed kinetic energy density, in the case where
$h_d/z_d=0.2$ with
 $\rho_{d0}(0)/\rho_g=2$,  $T=1$ and $\log(k_y^2 \eta^2 r^2)=3.55$.
The position of arrow  shows the stabilization time $t_s \Omega_K$.}
\label{fig:hdpe}
\end{figure}

\begin{figure}[p]
\centerline{
\epsfysize=12cm
\epsfbox{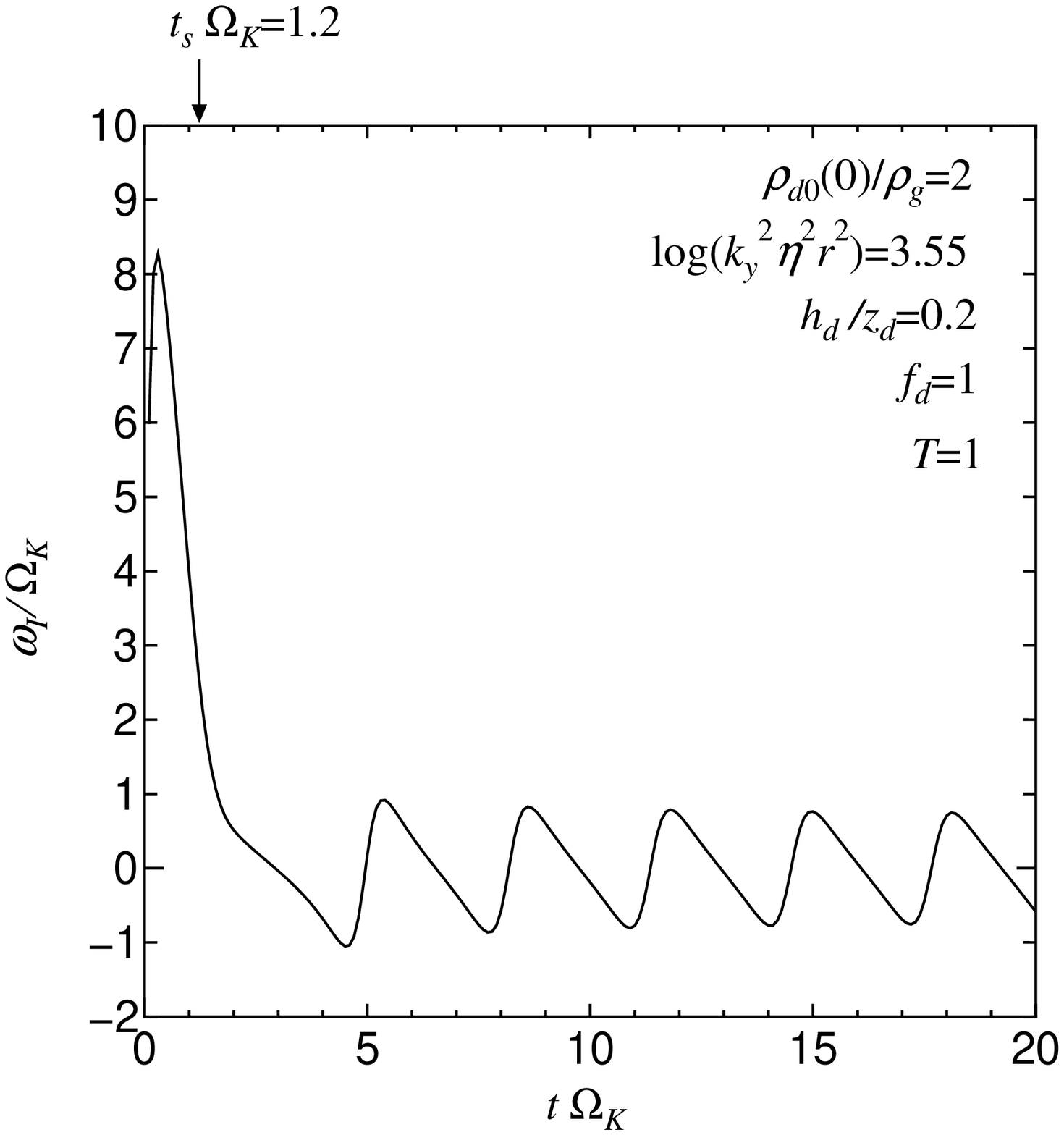}
}
\caption{The time evolution of  the growth rate, in the case where
$h_d/z_d=0.2$ with
 $\rho_{d0}(0)/\rho_g=1$,  $T=1$ and $\log(k_y^2 \eta^2 r^2)=3.55$.
The position of arrow  shows the stabilization time $t_s \Omega_K$.}
\label{fig:hdgr}
\end{figure}

Next, we consider the case where  $\rho_{d0}(0)/\rho_g=2$. The results
of the previous section show that the peak growth  rate of 
the instability for $\rho_{d0}(0)/\rho_g=2$ is much larger than 
the case $\rho_{d0}(0)/\rho_g=0.5$ as long as the effect of the tidal 
force is not taken into account, {\it i.e.} $T=0$ (see  Fig. \ref{fig:dcfig2}).
Figures \ref{fig:hdpe} and  \ref{fig:hdgr} show the time evolutions of 
components of energy and the growth rate, respectively.
Reflecting analytical results in the case of $T=0$, 
there is a steep rise in the early stage. 
The growth rate  decays during about a Kepler period after the early increase.
Although the instability  grows several orders of magnitude before 
stabilization (Fig. \ref{fig:hdpe}), we cannot definitely claim
whether the non-linear effect should work or not, since the initial 
amplitudes are unknown.

\begin{figure}[p]
\centerline{
\epsfysize=12cm
\epsfbox{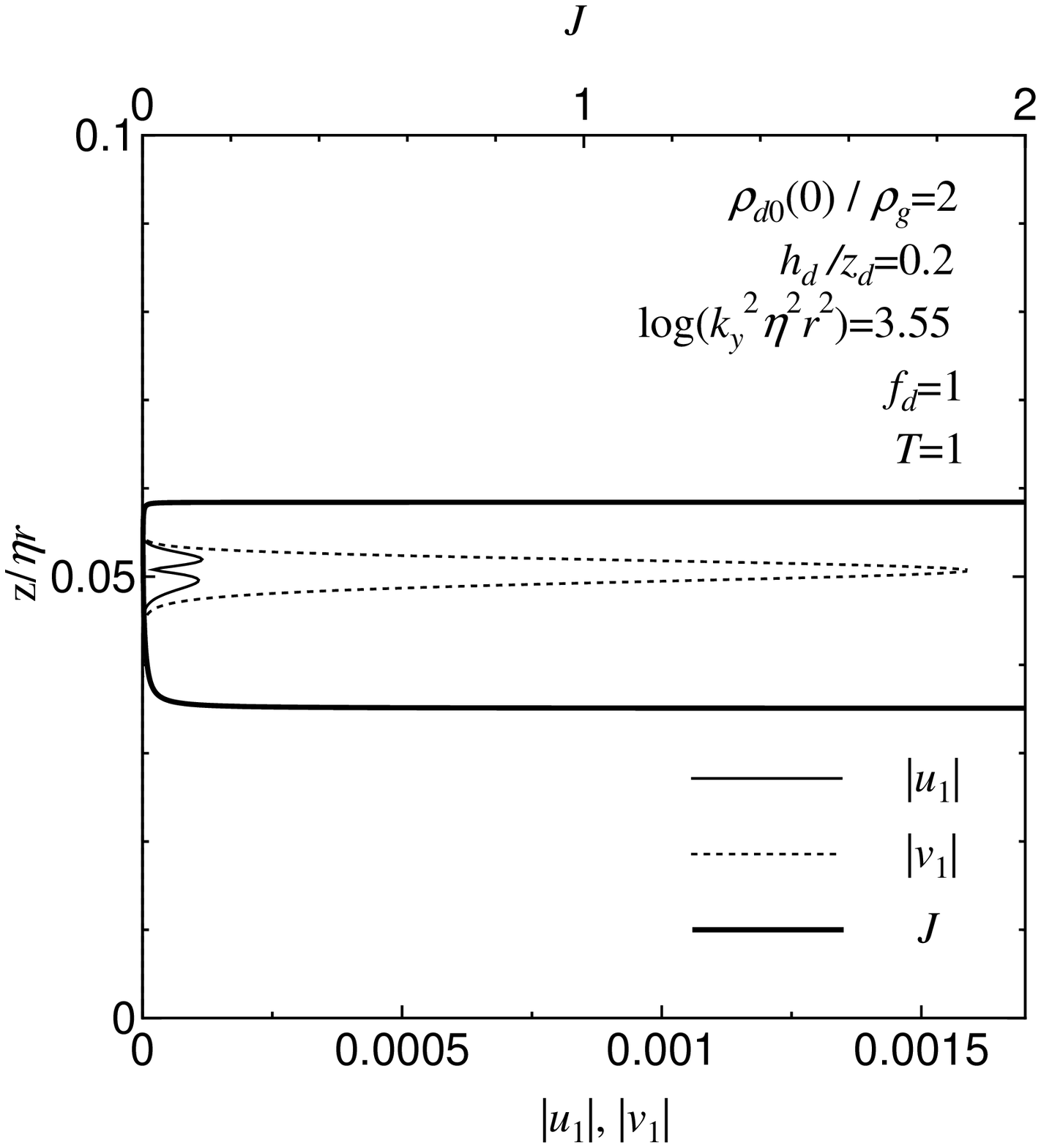}
}
\caption{Eigenfunctions of  the perturbed radial (solid curve) 
and azimutal velocity (dotted curve)
at a time when
 $<u_1^2>/2$ reaches a minimum value and 
 $<v_1^2>/2$ reaches a maximum value ( $t \Omega_K = 25.6$)
 in the case where
 $\rho_{d0}(0)/\rho_g=1$, $h_d/z_d=0.2$, $f_g=1$,  $f_d=1$, $T=1$
and  $\log(k_y^2 \eta^2 r^2)=3.55$.
A Bold solid curve denotes the Richardson number distribution.}
\label{fig:ef.umivmxnr1h1}
\end{figure}

\begin{figure}[p]
\centerline{
\epsfysize=12cm
\epsfbox{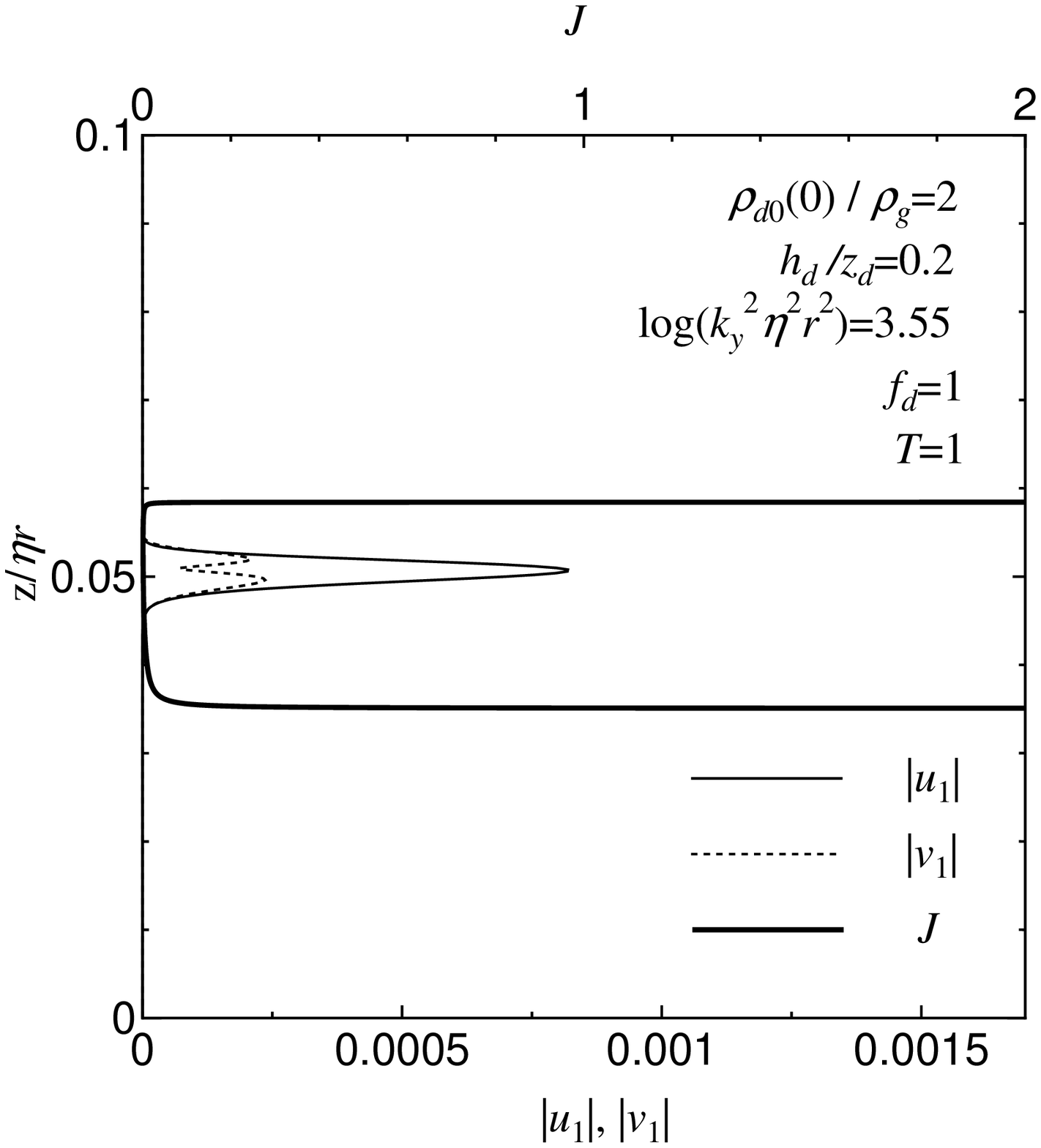}
}
\caption{Same as Fig. \protect\ref{fig:ef.umivmxnr1h1} except for
a time when  $<u_1^2>/2$ reaches a maximum value and 
 $<v_1^2>/2$ reaches a minimum value ( $t \Omega_K = 27.1$).}
\label{fig:ef.umxvminr1h1}
\end{figure}

Figures \ref{fig:ef.umivmxnr1h1} and \ref{fig:ef.umxvminr1h1} show 
absolute values of 
eigenfunction for $\rho_{d0}(0)/\rho_g=2$, $h_d/z_d=0.2$, $f_d=1$,
$T=1$  and $\log(k_y^2 \eta^2 r^2)=3.55$
when the radial kinetic energy  the smallest ($t \Omega_K = 25.6$)
and the largest ($t \Omega_K = 27.1$) values, respectively.
A bold solid curve denotes the Richardson number distribution.
The vertical velocity $|w_1|$ is omitted because it is of two orders smaller 
than $|u_1|$ and $|v_1|$ (see Fig \ref{fig:hdpe}).
The eigenfunctions have the largest amplitude at the height 
where the Richardson number has the  smallest value.

\begin{figure}[p]
\centerline{
\epsfysize=12cm
\epsfbox{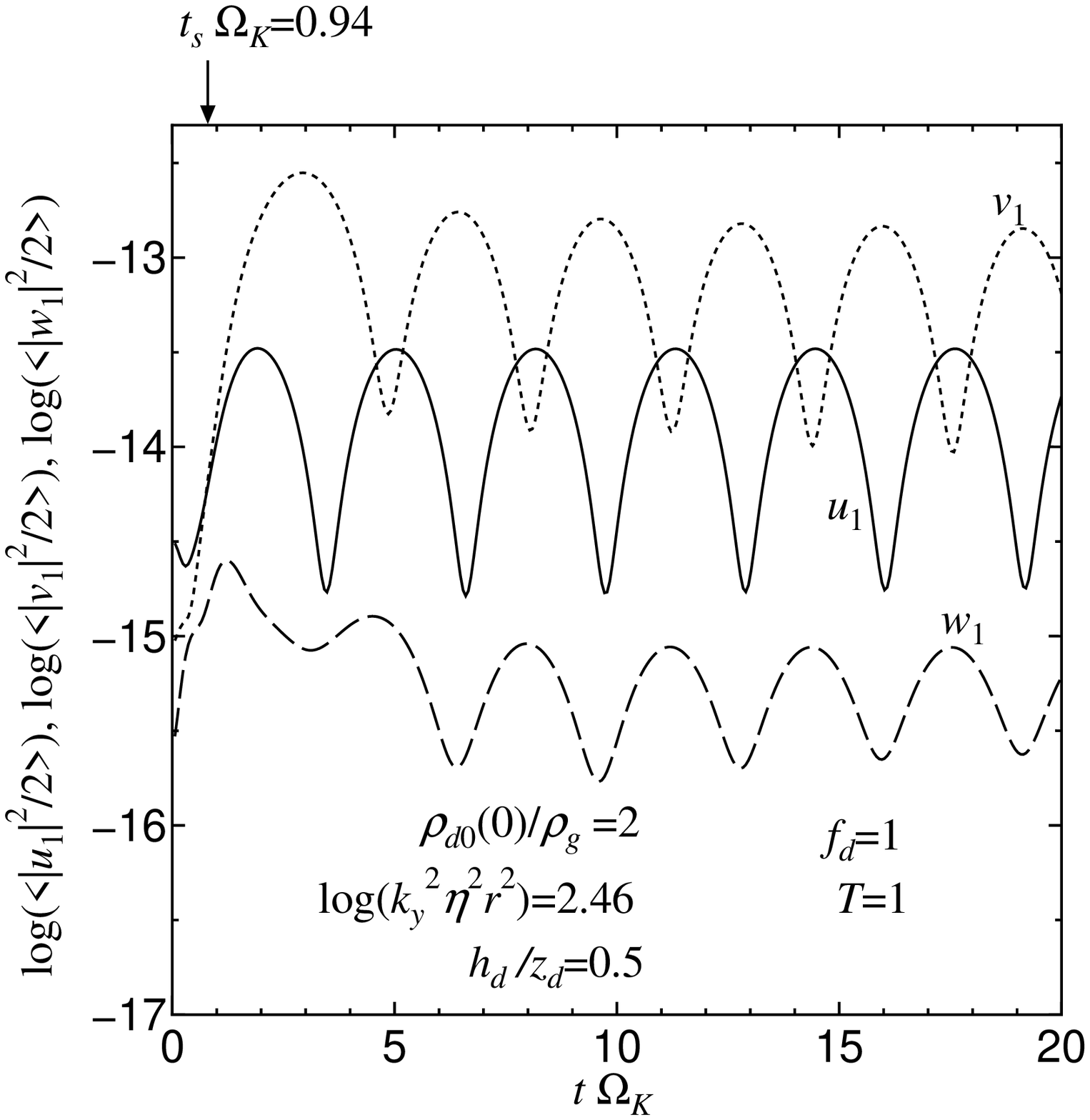}
}
\caption {The time evolution of  radial, azimuthal and vertical
parts of the perturbed kinetic energy density, in the case where 
$h_d/z_d=0.5$ with
 $\rho_{d0}(0)/\rho_g=2$, $T=1$ and $\log(k_y^2 \eta^2 r^2)=2.46$.
The position of arrow  shows the stabilization time $t_s \Omega_K$.}
\label{fig:er1h5}
\end{figure}

\begin{figure}[p]
\centerline{
\epsfysize=12cm
\epsfbox{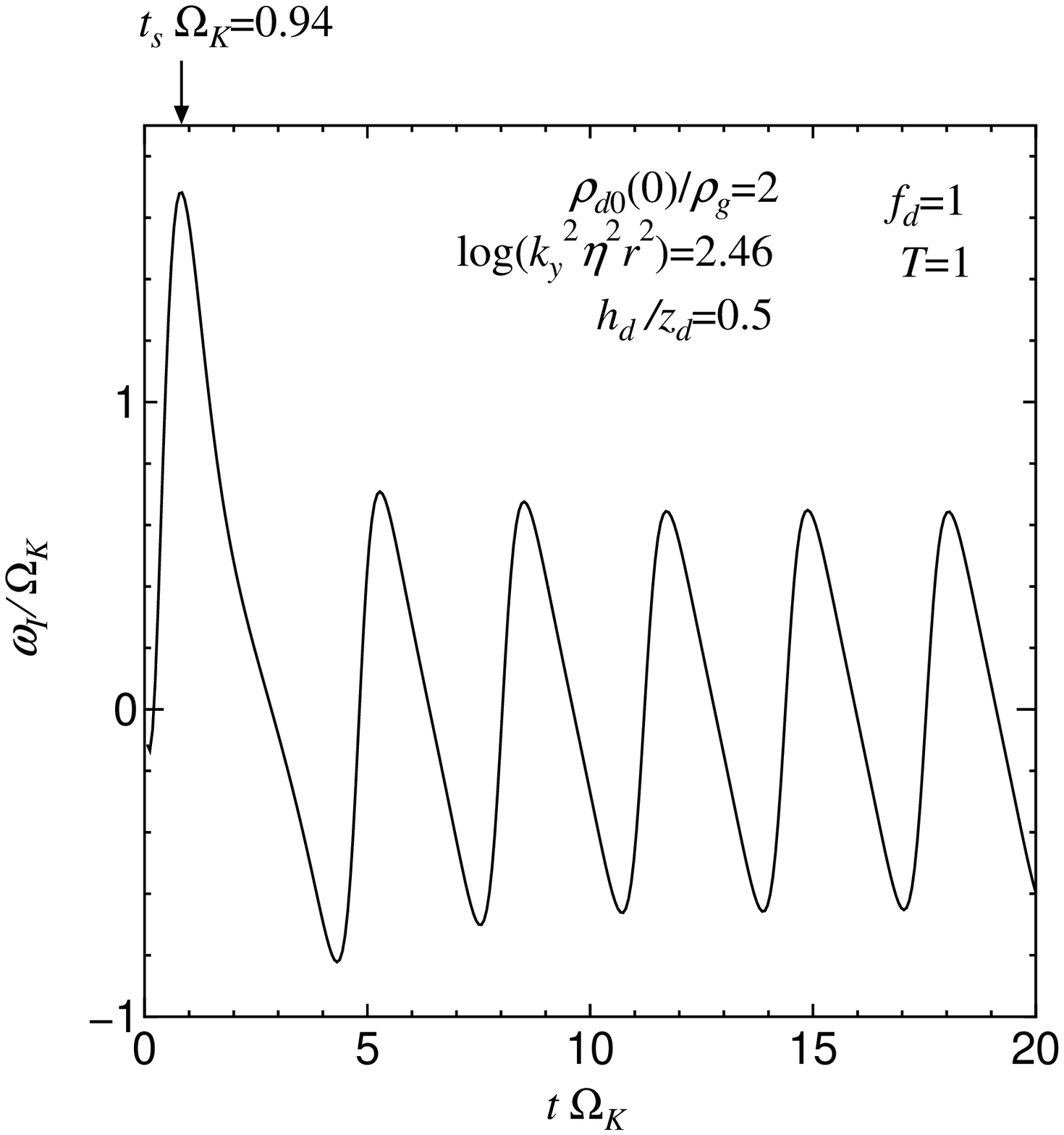}
}
\caption{ The time evolution of  the growth rate, in the case where
$h_d/z_d=0.5$ with
 $\rho_{d0}(0)/\rho_g=2$, $T=1$ and $\log(k_y^2 \eta^2 r^2)=2.46$.
The position of arrow  shows the stabilization time $t_s \Omega_K$.}
\label{fig:gr1h5}
\end{figure}

From Fig. \ref{fig:dcfig4} for $T=0$, a small initial growth rate 
of the instability is also expected for a large value of $h_d/z_d$.
Figures \ref{fig:er1h5} and \ref{fig:gr1h5} show the time evolutions 
of components of energy and the growth rate for 
$h_d/z_d=0.5$, $\log(k_y^2\eta^2 r^2)=2.46$, $T=1$ and 
$\rho_{d0}(0) /\rho_g= 2$,  respectively. 
This growth rate rises  at first and then begins to vibrate,
like the case where  $\rho_{d0}(0)/\rho_g=0.5$ and $ h_d/z_d=0.2$.

\begin{figure}[p]
\centerline{
\epsfysize=12cm
\epsfbox{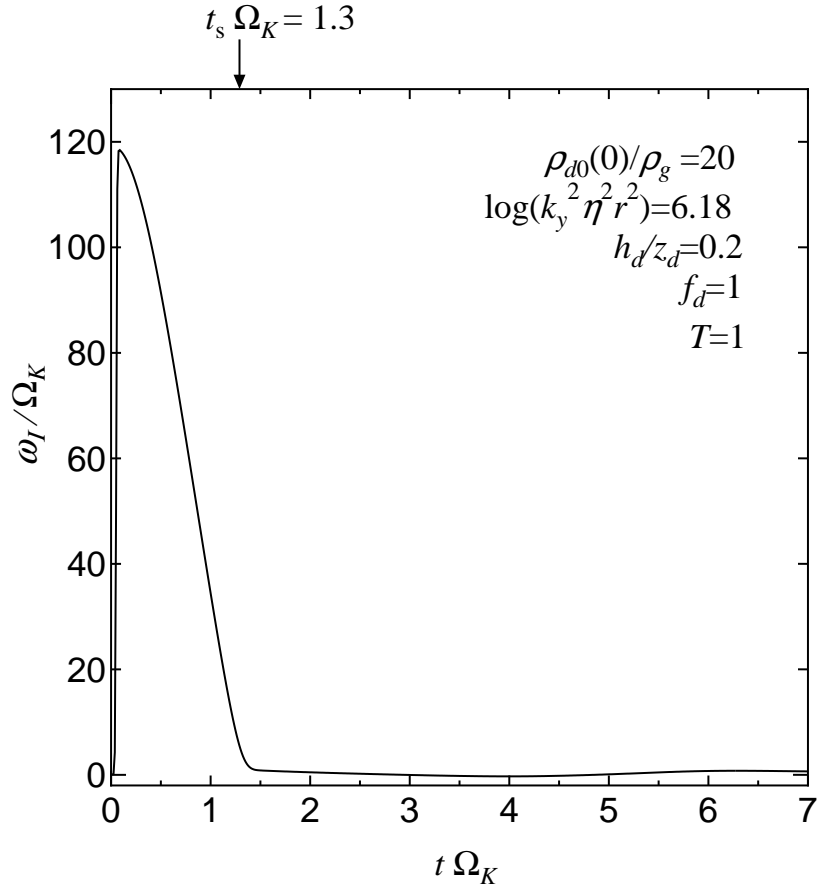}
}
\caption{The time evolution of  the growth rate, in the case where
$\log(k_y^2 \eta^2 r^2)=5.96$ 
with  $T=1, 
\rho_{d0}(0)/\rho_g=20$ and $h_d/z_d=0.2$.
The position of a arrow  shows the stabilization time $t_s \Omega_K$.}
\label{fig:tfig9}
\end{figure}

\begin{figure}[p]
\centerline{
\epsfysize=12cm
\epsfbox{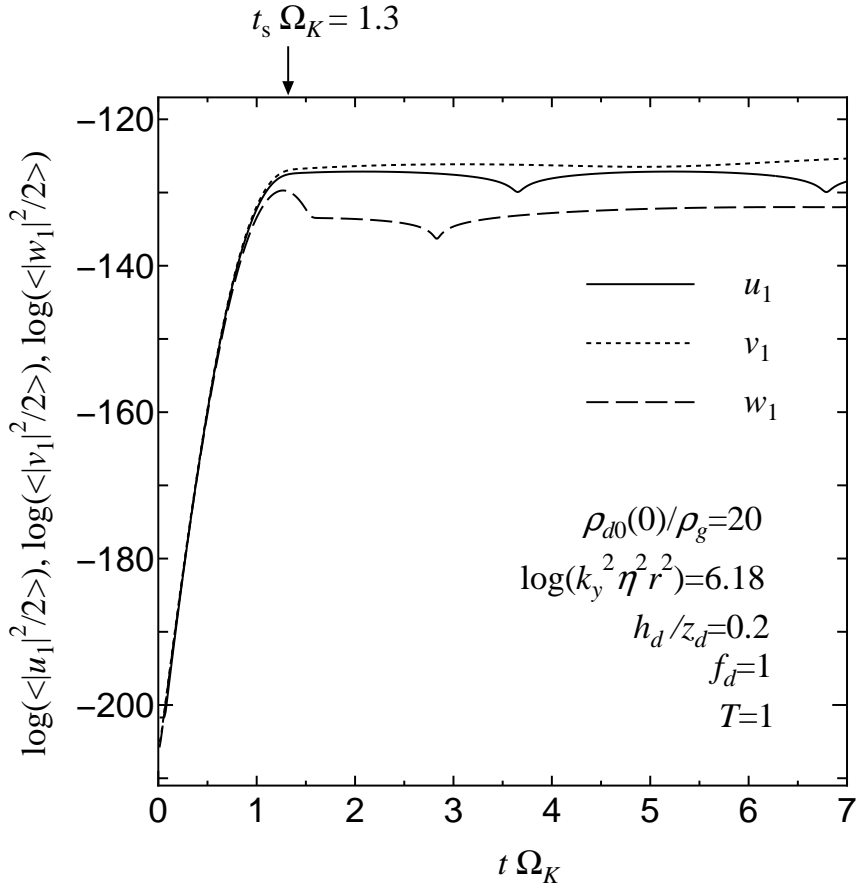}
}
\caption{The time evolution of  radial, azimuthal and vertical
parts of the perturbed kinetic energy density, in the case where
 $\log(k_y^2 \eta^2 r^2)=5.96$, with  $T=1$, 
$\rho_{d0}(0)/\rho_g=20$ and $h_d/z_d=0.2$.
The position of arrow  shows the stabilization time $t_s \Omega_K$.}
\label{fig:tfig10}
\end{figure}

\begin{figure}[p]
\centerline{
\epsfysize=12cm
\epsfbox{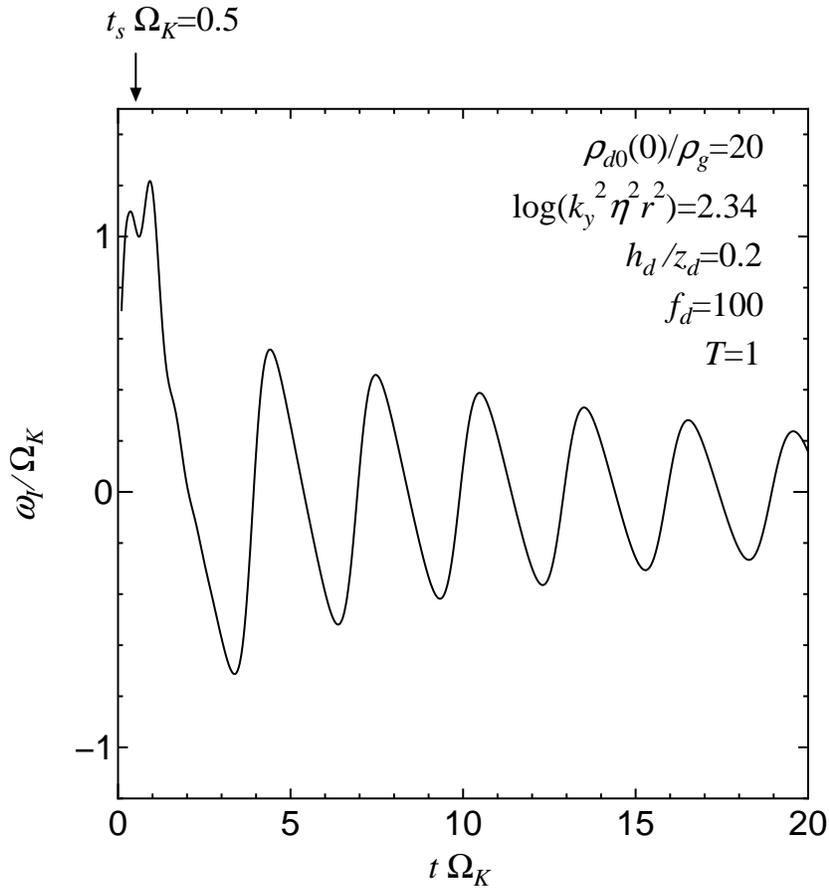}
}
\caption{ The time evolution of  the growth rate, in the case where
 $f_d=100$ and $\rho_{d0}(0)/\rho_g=20$, with $\log(k_y^2 \eta^2 r^2)=2.34$
and $h_d/z_d=0.2$.
The position of arrow  shows the stabilization time $t_s \Omega_K$. }
\label{fig:tfig15}
\end{figure}

\begin{figure}[p]
\centerline{
\epsfysize=12cm
\epsfbox{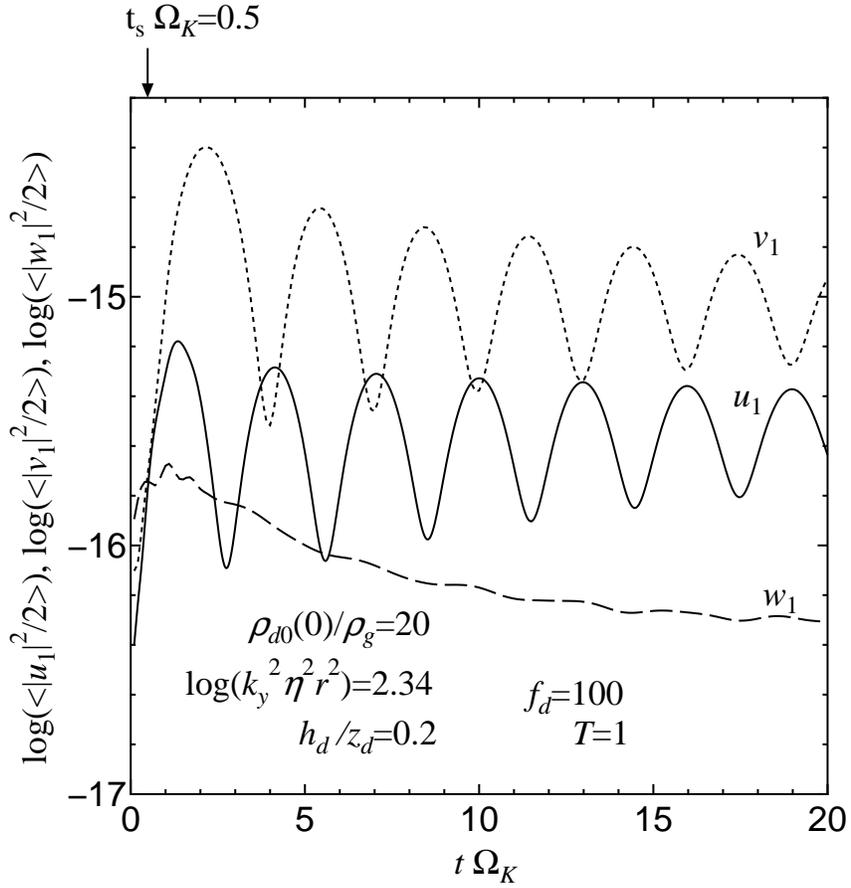}
}
\caption{The time evolution of  radial, azimuthal and vertical
parts of the perturbed kinetic energy density, in the case where
 $f_d=100$ and $\rho_{d0}(0)/\rho_g=20$, with $\log(k_y^2 \eta^2 r^2)=2.34$
and $h_d/z_d=0.2$.
The position of arrow  shows the stabilization time $t_s \Omega_K$.}
\label{fig:tfig16}
\end{figure}

From  Fig. \ref{fig:dcfig2}, the growth rate in the case $T=0$
is obviously  very large when dust settling proceeds 
and $\rho_{d0}(0)/\rho_g \geq 20$.
We calculated numerically the time evolutions in the cases 
 $\log(k_y^2 \eta^2 r^2)=6.18$  with $T=1$, $\rho_{d0}(0)/\rho_g=20$ 
and $h_d/z_d=0.2$ (Fig. \ref{fig:tfig9}).
When $\log(k_y^2 \eta^2 r^2)=6.18$, we found 
$k_{xc} = 2.4 \times 10^3$ for $T=0$, and then
we have  $t_s \Omega_K = 1.3$ from Eq. (\ref{eqn:crit}).
The estimated value of the time for stabilization is displayed
 by the arrow in Fig. \ref{fig:tfig9}, which is consistent 
with the stabilization time of the numerical solution. 
During several Keplerian periods,
the perturbed kinetic energy increases by more than 100 orders for
$\rho_{d0}(0)/\rho_g=20$ in this  linear  calculation in the case
 $\log(k_y^2 \eta^2 r^2)=6.18$ (Fig. \ref{fig:tfig10}).

Our results indicate that the perturbation increases
sufficiently, so that the nonlinear effect would be important, 
before the mode is stabilized. Thus, the tidal force is effective only in 
early phase of the dust settling where $\rho_{d0}(0)/\rho_g \simle 1$. 
The shear instability eventually grows to nonlinear phase 
when $\rho_d(0)/\rho_g$ becomes on the order of $10$ by the dust 
settling as long as $f_d/f_g=1$ ($i.e.$ the dust-gas surface density 
ratio is given by the solar elemental abundance).

Last, we consider the case where $\rho_{d0}(0)/\rho_g = 20$ and 
$f_d=100$ to see the dependency of the shear instability 
on $f_d$, that is, the dust surface density 
enhancement relative to that the solar abundance
(see Figs. \ref{fig:tfig15} and \ref{fig:tfig16}). 
The shear instability is stabilized 
before the perturbed kinetic energy grows largely in this case.
According to  results without the tidal force, the peak 
growth rate in this case is about a Kepler frequency 
and  comparable  to that  in the case with
$\rho_{d0}(0)/\rho_g \approx 0.5$ and $f_d=1$ (compare Fig. \ref{fig:dcfig2}
 and \ref{fig:fd}).
Thus, even though $\rho_{d0}(0)/\rho_g$ increases, 
the shear instability is  stabilized
and planetesimals can be formed by the gravitational instability 
if dust concentrates to some regions in the solar nebula.

As stated in \S IV, decrease of $f_g$ is idntical to increase of $f_d$.
This implies that the local dissipation of gas also causes 
stabilization of the shear instability.

\section{CONCLUSIONS AND DISCUSSION}

The linear stability analysis of the dust layer in 
the solar nebula is done 
including the effects of both the Coriolis and the tidal forces.  
The hybrid density distribution of the dust is assumed 
where the dust density is constant in the dust layer 
and the transition region has a sinusoidal density distribution.
The calculations with both the Coriolis and the tidal forces show 
that the tidal force has a stabilizing effect.  
The tidal force causes the radial shear in the disk.  
This radial shear changes the radial wave number 
of the mode which is at first unstable, 
and the mode is eventually stabilized.  
Thus the behavior of a mode is divided into two stages:
(1) the first growth of the unstable mode 
which is similar to the results without the tidal force, and
(2) the subsequent stabilization due to an increase of the radial wave number 
by the radial shear.  
If $\rho_d(0)/\rho_g \simle 2$, the stabilization occurs before 
the unstable mode grows largely.
On the other hand, the mode grows more than hundreds orders of magnitude, 
if $\rho_d(0)/\rho_g \simge 20$.
 
Since $\rho_c/\rho_g$ is much larger than 20,
 the linear hydrodynamic instability 
calculated in this paper grows largely before dust settling proceeds 
enough for the dust layer to be gravitationally unstable.
 If this instability develops turbulence 
in the transition region between the dust layer and the gas layer, 
the overshoot of the turbulence would widen the transition region, 
and the transition region would be stabilized, 
because the growth rate of the shear instability decreases 
as the half-thickness of the transition region increases.  
Then, the dust aggregates settle further, and the transition region becomes unstable again.   
Repeats of this cycle would eventually lead the whole of the dust layer to be a quasi-equilibrium state where the dust settling and turbulent diffusion balance each other.
 Note that this equilibrium state is not necessarily given by $J$=constant distribution derived by Sekiya (1998), because the constant $J$ distribution is derived by neglecting the effects of the Coriolis and tidal forces.  
If this equilibrium state is realized before the density exceeds the critical density of the gravitational instability, 
planetesimals cannot be formed by the gravitational instability.  
Further nonlinear calculations on the evolution of the dust layer 
should be done in the future in order to see whether turbulence really develops and an equilibrium state is realized.   

We have assumed that the surface density distribution is given by 
the Hayashi model ({\it i.e.} $f_d =1$, $f_g=1$)
 for most parts of calculations in this paper.
Our results show that the shear instability is stabilized 
by the tidal force if the ratio of the dust surface density $\Sigma_d$ 
to the gas surface density $\Sigma_g$ is hundreds times as large 
as that calculated from the solar elemental abundance,
even if $\rho_{d0}(0) \sim \rho_c$.
The shear-induced instability does not grow largely 
when the dust density reaches the critical density of the 
gravitational instability in this linear calculation
if dust concentrates to some locations, or gas dissipates from the nebula.  
For example, dust concentration due to radial migration induced 
by the gas drag in the laminar disk could increase 
the dust to gas mass ratio (Youdin and Shu 2002). 
Then the planetesimal formation due to gravitational instability may occur.  
In this case, the effect of the self-gravity, e.g. midplane cusp 
of the vertical density distribution (Sekiya 1998, Youdin and Shu 2002) 
should be taken into account. 

Lastly, we discuss the dust concentration by eddies in a protoplanetary disk.  
Cuzzi {\it et al.} (2001) assumed a turbulent disk 
and showed that the dust concentrates to preferential positions between eddies, 
even if the dust aggregates are as small as 1mm.  
In their analysis, the effects of the solar gravity are not included.  
If dust concentrated to some positions, they revolve around the central star 
with the Keplerian velocity.  
On the other hand, the residual gas rich parts revolve with smaller mean velocity 
due to the radial pressure gradient.  
Thus, the dust aggregates may be removed from the preferential positions between eddies 
and the concentration may be suppressed.  
There are also models of dust concentration by large eddies.  
Klahr and Henning (1997) assumed large scale eddies whose rotation axis are parallel to the midplane.  
They showed that dust concentration away from the midplane.  
On the other hand, another ``vortex capture" model 
(Barge and Sommeria 1995, Tanga {\it et al.} 1996, Godon and Livio 1999, 2000,
Chavanis 2000, de la Fuente Marcos and Barge 2001) 
assumes eddies whose rotational axis is parallel to the rotation axis of the protoplanetary disk.  
These eddies preferentially concentrate dust aggregates 
which have stopping time due to gas drag on the order of the Keplerian period.  
These models give an additional possibility to reach the gravitational instability.  
However, it is not clear what kind of eddies are really probable in the protoplanetary disks.   
Full hydrodynamic simulation including interactions between gas and dust and the solar gravity 
should be performed in future to elucidate whether 
the dust concentration by eddies really occurs in the protoplanetary disks.

\section*{ACKNOWLEDGMENTS}
We thank  Drs. Palo Tanga, Andrew Youdin, Sei-Ichiro Watanabe,
 Saburo Miyahara and Shin-Ichi Takehiro for valuable comments.
Numerical calculations were performed
partly at the Astronomical Data Analysis Center
of the National Astronomical Observatory, Japan.

\newpage

\begin{center}
Table I \\
Model and Numerical Parameters \\
~\\
\begin{tabular}{ccccccccccc} \hline
$\log(k_y^2 \eta^2 r^2)$ & $k_x \eta r$ & 
$\rho_{d0}(0) /\rho_g $ & $h_d /z_d $& $f_g$ & $f_d$ & $T$ & $J_{min}$&
$N_z$ & $z_0 / \eta r$ & $\Delta t ~ \Omega_K$ \\ \hline \hline 
 2.46 & 0 & 0.5 & 0.2 &1 & 1 & 0 & $3.94\times 10^{-2}$ & 2048 & 0.8 
& $5 \times 10^{-3}$ \\ %\hline
 2.46 & 0 & 0.5 & 0.2 &1 & 1 & 1 & $3.94\times 10^{-2}$ & 2048 & 0.8 
& $5 \times 10^{-3}$ \\ %\hline
 2.00 & 0 & 0.5 & 0.2 & 1 & 1 & 1 & $3.94\times 10^{-2}$ & 2048 & 0.8
 & $5 \times 10^{-3}$ \\ %\hline
 2.46 & 0 & 0.5 & 0.2 & 1 &1 & 0.5 & $3.94\times 10^{-2}$ & 2048 & 0.8
 & $5 \times 10^{-3}$ \\ % \hline
 2.46 & 0 & 0.5 & 0.2 & 1 &1 & 0.1 &  $3.94\times 10^{-2}$ & 2048 & 0.8 
& $5 \times 10^{-3}$ \\ % \hline
 2.46 & 50 & 0.5 & 0.2 & 1 &1 & 0 & $3.94\times 10^{-2}$ & 2048 & 0.8
& $5 \times 10^{-3}$ \\ % \hline
 3.55 & 0 & 2 & 0.2 & 1 &1 & 1 &  $1.15\times 10^{-3}$ & 8192 & 0.2
& $5 \times 10^{-4}$ \\ %\hline
 2.46 & 0 & 2 & 0.5 & 1 &1 & 1 &$ 3.35\times 10^{-3}$ & 2048 & 0.5 
& $5 \times 10^{-3}$ \\ %\hline
 6.18 & 0 & 20 & 0.2 & 1 &1 & 1 & $6.33\times 10^{-6}$ & 65536 & 0.05
& $1 \times 10^{-5}$ \\ %\hline
 2.34 & 0 & 20 & 0.2 & 1 & 100  & 1 & $3.68\times 10^{-2}$ &  4096 & 1.0
& $5 \times 10^{-3}$ \\ \hline 

\end{tabular}
\end{center}	

\newpage

%Figure.1

%Figure.2

%Figure.3

%Figure.4

%Figure.5

%Figure.6

%Figure.7

%Figure.8

%Figure.9

%Figure.10

%Figure.11

%Figure.12

%Figure.13

%Eigenfuction
\clearpage

%Figure.14

%Figure.15

%Figure.16

%Figure.17

%Figure.18

%Figure.19

%Figure.20

%Figure.21

\end{document}